\documentclass[pre,aps,reprint,longbibliography,amsmath,amsfonts,amssymb,twocolumn,showpacs]{revtex4-1}
\usepackage{CJK,graphicx,bm,xspace}
\newcommand{\miAPT}{miAPT\xspace}
\newcommand{\maAPT}{maAPT\xspace}
\newcommand{\A}{\mathcal{A}}
\newcommand{\mI}{\mathcal{I}}

\begin{document}
\begin{CJK*}{UTF8}{mj}
\title{Absorbing phase transitions in deterministic fixed-energy sandpile models}
\author{Su-Chan Park (박수찬)}
\affiliation{Department of Physics, The Catholic University of Korea, Bucheon 14662, Republic of Korea}
\date{\today}
\begin{abstract}
We investigate the origin of the difference, which was
	noticed by Fey {\it et al.} [Phys. Rev. Lett. {\bf 104}, 145703 (2010)], between the 
steady state density of an Abelian sandpile model (ASM) 
and the transition point of its corresponding deterministic 
fixed-energy sandpile model (DFES). 
Being deterministic, the configuration space of a DFES 
can be divided into two disjoint classes such that  
every configuration in one class should evolve into one of absorbing states,
whereas no configurations in the other class can reach an absorbing state.
Since the two classes are separated in terms of toppling dynamics,
	the system can be made to exhibit an absorbing phase transition (APT) at 
various points that depend on the initial probability distribution of 
the configurations. 
Furthermore, we show that in general the transition point also depends on whether
an infinite-size limit is taken before or after the infinite-time limit.
To demonstrate, we numerically study the two-dimensional DFES with 
Bak-Tang-Wiesenfeld toppling rule (BTW-FES).
We confirm that there are indeed many thresholds.  
Nonetheless, the critical phenomena
at various transition points are found to be universal.
We furthermore discuss a microscopic absorbing phase 
transition, or a so-called spreading dynamics, of the BTW-FES, to find that the phase transition in this setting is related to the dynamical 
isotropic percolation process rather than self-organized criticality. 
In particular, we argue that choosing recurrent configurations of the 
corresponding ASM as an initial configuration does not allow for
a nontrivial APT in the DFES.
\end{abstract}
\maketitle
\end{CJK*}
\section{\label{Sec:intro}Introduction}
A fixed-energy ensemble~\cite{TB1988,DVZ1998,VDMZ1998,VDMZ2000} 
of sandpile models~\cite{BTW1987,D1990,M1991,IP1998}
was introduced to interpret
self-organized criticality (SOC)~\cite{BTW1987,J1998book,P2012book} in the 
context of the absorbing phase transition (APT) (for a review, see, e.g.,
Refs.~\cite{H2000,O2004}).
The idea of introducing the fixed-energy sandpile model (FES) stems from 
the similarity between
a stable configuration of the sandpile model 
and an absorbing state. By an absorbing state we mean
a configuration that does not change with time by the given dynamic rules.
In a sense, it was believed that the FES is to the 
sandpile model as the canonical ensemble is to the grand canonical 
ensemble in equilibrium statistical mechanics. 
By the same token, a phase transition in a FES was expected to occur 
at the steady state density of the corresponding sandpile model.

However, it was noticed~\cite{FLW2010} that for a certain deterministic
fixed-energy sandpile model (DFES), the critical energy density
is different from the steady state energy density
of the corresponding Abelian sandpile model (ASM)~\cite{D1990}.
This observation clearly raises many questions.
Why do some DFESs have the same critical density but some do not?
What is the origin of the difference? 
Is the critical behavior of the FES at different transition points universal?
Can we expect similar differences in the stochastic fixed-energy
sandpile model (SFES)? 
The aim of this paper is to answer these questions.

This paper is organized as follows. In Sec.~\ref{Sec:DFES},
we define the DFES in order to introduce the notations
we are using in this paper. In Sec.~\ref{Sec:MAPT}, 
we discuss the APT in the DFES.
We show that in general
the transition points are dependent on the initial condition. 
Furthermore, we show that transition points can vary,
according to which limit is taken first, the infinite-size limit
or the infinite-time limit.
As a nontrivial example, Sec.~\ref{Sec:btwdfes} numerically
investigates APTs in the
Bak-Tang-Wiesenfeld (BTW)-type FES (BTW-FES) in two dimensions.
Indeed, the critical point of the BTW-FES is found to vary
with the initial condition. We also show that the critical
behavior is universal.
In Sec.~\ref{Sec:Dis}, we discuss the implication of our observation
to the SOC models and the SFES.
In Sec.~\ref{Sec:sum}, we summarize the work.

\section{\label{Sec:DFES}Model}
In this section, we define the DFES to introduce the notations in this paper.
We consider a system with $V$ sites. 
A configuration of the system is specified by 
a non-negative integer $z_i$ at each site $i$ ($1\le i \le V$), which 
is called the \emph{energy} at site $i$.
The time evolution of the system is governed by a $V\times V$ integer matrix $F$, 
to be called a toppling matrix (TM), with the following properties:
\begin{enumerate}
\item $F_{ij} \le 0$, if $i \neq j$,
\item $F_{ii} = -\sum_{j\neq i} F_{ij}$.
\end{enumerate}
A TM is said to be irreducible if,
for any ordered pair of two different site indices $(i,j)$ $(1\le i,j\le V)$, there is an ordered set of different indices $(k_1,k_2,\ldots,k_{n-1})$
such that 
$\prod_{l=0}^{n-1} F_{k_l k_{l+1}}  \neq 0$,
where $k_0 \equiv i$ and $k_n \equiv j$. 
Obviously, $F_{ii}>0$ for all $i$ if $F$ is irreducible.
In this paper, a TM is always assumed irreducible.
For a given  TM, we will say that site $i$ is active if $z_i \ge F_{ii}$. 
An active site $i$ will topple and distribute its energy to other sites
in such a manner that $z_j \mapsto z_j - F_{ij}$ for all $j$.

We consider two rules of time evolution.
One is the parallel update rule (PU), and the other is
the random sequential update rule (RSU).
In the PU, all active sites at time $t$ topple simultaneously and
the configuration at time $t+1$ becomes ($i=1,\ldots,V$)
\begin{align}
z_i\left (t+1\right ) = z_i\left ( t \right ) - \sum_{n=1}^N F_{j_n i},
\end{align}
where $j_n$'s are the indices of the active sites at time $t$ 
and we have assumed that there are $N$ active sites.
If there are no active sites, the system does not change with time.
In this sense, any configuration without active sites is an absorbing state.

In the PU, 
the system evolves deterministically and time takes integer values.
In the RSU, on the other hand, the system evolves stochastically and time takes continuous 
values. Assume that there are $N_t$ active sites at time $t$.
One of the active sites is randomly chosen with equal probability,
and energy at every site $j$ at time $t+dt$ becomes
\begin{align}
z_j \left (t+dt\right ) = z_j\left (t\right ) - F_{ij},
\end{align}
where $dt$ is an exponentially distributed random number with mean $1/N_t$ and
$i$ is the index of the chosen active site.
As in the PU, the system does not change with time if it is in an absorbing state.

The RSU actually simulates the master equation 
\begin{align}
\frac{\partial}{\partial t} P(C,t) 
= \sum_{j} P(C_j,t) -N(C) P(C,t),
\label{Eq:master}
\end{align}
where $P(C,t)$ is the probability of being in configuration $C$ at time $t$, 
$N(C)$ is the number of active sites in configuration $C$,
and $C_j$ is the configuration constructed by adding $F_{ji}$ to $z_i$ for all $i$ in configuration $C$. Note that $P(C_j,t)=0$ if $C_j$ has a site with 
negative energy.

Due to the second property of a TM, the total energy defined as
\begin{align}
E \equiv \sum_{i} z_i
\label{Eq:Energy}
\end{align}
does not change with time.
When the total energy is low, it is likely that the system will eventually
fall into one of the absorbing states. Meanwhile, if the total energy is
high, the system may change with time indefinitely.
Hence, as the total energy varies, the system can 
exhibit an APT.
In the next section, we will discuss the general features of the APT 
exhibited by the DFES.

\section{\label{Sec:MAPT}Absorbing phase transitions: general features}
If there is nonzero probability current going into
an absorbing state from any configuration, 
a system with a finite configuration space will eventually 
fall into an absorbing state (for a proof, see, e.g., Ref.~\cite{vK1992}). 
Thus, loosely speaking, a sharp phase transition can be defined only if 
there are configurations with zero probability current going into an absorbing state.
For instance, in the contact process~\cite{H1974}, which is a prototypical
model of an APT, 
such configurations can emerge only under the infinite-size limit.
Accordingly, the infinite-size limit should be taken before the infinite-time
limit to study a nontrivial APT of the contact process. 
Actually, this order of the two limits is generally taken for a phase transition of a dynamic system.

As we will see soon, however, the DFES is somewhat exceptional in that 
a finite system can have configurations with zero probability current going into 
any of absorbing states. 
Hence, a nontrivial transition can exist in a finite system.
Even in the case that a phase transition 
is well defined only in the infinite-size limit (see below),
it can exist regardless of whether the infinite-size limit
is taken before or after the infinite-time limit. 
However, the definition of an order parameter that is used to locate a transition 
point should depend on which limit is taken first. 
In the following, we define three different order parameters
and discuss the general features of a phase transition 
associated with each one.
\subsection{\label{Sec:phi1} Order parameter $\phi_1$}
Due to the deterministic nature of the PU, the fate 
of a finite system (to survive or not to survive) 
is already determined from the outset.
The RSU cannot change the fate because of
the Abelian property of toppling~\cite{D1990}. On this account, we can 
define, regardless of the update rules, the survival function $S$ of 
configurations as
\begin{align}
S(C) = \lim_{t\rightarrow \infty} \mathrm{Pr}_a(t|C),
\label{Eq:tildeS}
\end{align}
where $\mathrm{Pr}_a(t|C)$ is the probability that the system at time $t$ has active sites 
if it starts from configuration $C$ at $t=0$ 
(under the PU, $\mathrm{Pr}_a$ is either 1 or 0).
Note that the infinite-size limit is not assumed in Eq.~\eqref{Eq:tildeS}.

Since the possible value of $S$ is either one or zero, 
we can divide the set of all configurations by the two disjoint 
sets $\mI$ and $\A$,
\begin{align}
\mI = \{ C | S(C) = 0\},\quad
\A = \{ C | S(C) = 1\}.
\label{Eq:SIA}
\end{align}
In fact, both $\mI$ and $\A$ can be further divided into many disjoint classes 
according to the toppling invariant~\cite{Dhar2006} as well as total energy. 
For our purpose, however, it is enough to consider only these two classes.

Since any configuration with a total energy lower than $\text{min}\{F_{ii}\}$ cannot have an active site, $\mI$ is not empty
and the energy of a configuration in $\A$ cannot be lower than $\text{min}\{F_{ii}\}$.
Likewise, since any configuration with a total energy higher than 
$\sum_i (F_{ii}-1)$ cannot be absorbing, $\A$ is not empty and 
the energy of configurations in $\mI$ cannot be higher than $\sum_i (F_{ii}-1)$.
Thus, the total energy of configurations in $\mI$ ($\A$) is bounded from above (below).
Considering the existence of the bounds,
we define
\begin{align}
Z_M &= \sup\left \{ E(C) | C \in \mI \right \},\quad \zeta_M \equiv \frac{Z_M}{V},\nonumber\\
Z_m &= \inf\left \{ E(C) | C \in \A \right \},\quad \zeta_m \equiv \frac{Z_m}{V},
\end{align}
where, and in what follows, $E(C)$ means the total energy of configuration $C$.
Obviously, $Z_M = \sum_i (F_{ii}-1)$ for any TM (note that the configuration with $z_i = F_{ii}-1$ for all $i$ 
is an element of $\mI$). Although we could not find a general formula of $Z_m$ 
applicable to any TM, we found for a symmetric TM (see Appendix~\ref{App:Zm}),
\begin{align}
Z_m = \frac{1}{2} \sum_{i} F_{ii}.
\label{Eq:ZmSym}
\end{align}

Since any configuration with energy $Z_M + 1$ or higher always has an 
active site, $Z_m$ cannot be larger than $Z_M+1$. In Appendix~\ref{App:tree}, 
we show that for a symmetric TM the relation $Z_m = Z_M+1$ holds if and only if it is a tree
(for the definition of a tree, see Appendix~\ref{App:tree}).
In the following discussion,  we will mostly assume $Z_m \le Z_M$ and
the phase transition of a tree will be discussed only at the end of this section.

Let an initial configuration be randomly drawn from a probability distribution
$P_0(C;\zeta,V)$, where $V$ is the number of sites and
$\zeta$ is the (mean) energy density satisfying the self-consistent
condition
\begin{align}
\zeta = \frac{1}{V} \sum_{C} E(C) P_0(C;\zeta,V).
\label{Eq:defzeta}
\end{align}
Although the total energy of a configuration should be an integer,
$\zeta$ in Eq.~\eqref{Eq:defzeta} can take any positive real number.
When the infinite-size limit is involved, we will always assume
\begin{align}
\lim_{\Delta \rightarrow 0} \frac{1}{ \Delta}
\lim_{V\rightarrow \infty} \text{Prob}\left ( 
\left |\frac{E}{V} - e \right |< \frac{\Delta}{2} \right ) =
\delta(e-\zeta).
\label{Eq:Edis}
\end{align}

Now, we define
\begin{align}
\phi_1(\zeta;V) \equiv \sum_{C} S(C) P_0(C;\zeta,V),
\end{align}
which lies in the range $0 \le \phi_1 \le 1$.
If we take $\phi_1$ as an order parameter of an APT, as was implicitly
done in Ref.~\cite{FLW2010}, we can in fact make any value of $\zeta$ 
in the range $\zeta_m \le \zeta \le \zeta_M$  be a transition point 
by an appropriate choice of $P_0$.
To affirm this claim, we consider the initial distribution 
\begin{align}
P_0(C,\zeta;V) =& [1- f(\zeta)] P_I(C,\zeta)+ f(\zeta) P_A(C,\zeta),
\label{Eq:P0}
\end{align}
where $P_I$ and $P_A$ are any probability distributions with the restrictions
\begin{align} 
&P_I(C,\zeta)=0 \text{ if } C \in \A,\quad
P_A(C,\zeta)=0 \text{ if } C \in \mI,\nonumber \\
 &V\zeta=\sum_C E(C) P_I(C,\zeta) = \sum_C E(C) P_A(C,\zeta),
\end{align}
and $f(\zeta)$ is an arbitrary function only with the restriction $0 \le f(\zeta) \le 1$.
For any TM, we can in principle find $P_I$ and $P_A$ with the above properties 
as long as $\zeta_m \le \zeta \le \zeta_M$,
because there should be at least two configurations 
$C_1$ and $C_2$ such that $C_1 \in \mI, C_2 \in \A$, and $E(C_1) = E(C_2) = E$ 
for any (integer) energy in the range $Z_m \le E \le Z_M$.
With this initial condition, $\phi_1(\zeta) =f(\zeta)$ obviously.
Thus, by choosing appropriate $f$, we can have a phase transition at any point.
Even more, we can have any arbitrary value $\beta$ as a `critical' exponent of the order parameter $\phi_1$, by taking $f(\zeta) \sim |\zeta-\zeta_c|^\beta$ 
for $|\zeta-\zeta_c|\ll 1$ (in fact, the system at that transition point 
is not critical,
because the transition occurs in a finite system).

Since the system size is not involved in the above discussion, 
a phase transition can exist even for finite $V$. 
In case we want $\phi_1$ to show a sharp phase transition 
only in the infinite-size limit,
we only have to replace $f(\zeta)$ by a $V$-dependent function $g(\zeta,V)$ such that
$g$ is strictly positive for any finite $V$ and $\lim_{V\rightarrow \infty} g(\zeta,V) = f(\zeta)$.

We now use an example to support the above conclusion. 
Consider the TM ($1 \le i,j \le V$) 
\begin{align}
F_{ij} = 4 \delta_{ij} - 2 \delta_{|i-j|,1}-2 \delta_{|i-j|,V-1},
\label{Eq:BG}
\end{align}
which was called the bracelet graph in Ref.~\cite{FLW2010}.
According to the above discussion, we get $Z_m = 2V$, $Z_M = 3V$ or
$\zeta_m = 2$, $\zeta_M = 3$. Now we will give an initial condition yielding a transition point  in the range $2 \le \zeta_c \le 3$.

Since parity is conserved at all sites in any toppling event~\cite{FLW2010}, 
the number $w_i$ of pairs at each site $i$,  defined as
\begin{align}
w_i = \frac{1}{2} (z_i - b_i), 
\label{Eq:wi}
\end{align}
where $b_i = z_i \pmod{2}$, will determine the fate of the system.
Note that the dynamics of $w_i$ is identical to the DFES with the TM 
\begin{align}
\tilde F_{ij} = 2 \delta_{ij} - \delta_{|i-j|,1}- \delta_{|i-j|,V-1},
\label{Eq:BG_red}
\end{align}
which is exactly solvable~\cite{D2006}.
Since $Z_m = Z_M = V$ for $\tilde F$, $S(C)$ for $F$ is
\begin{align}
S(C) = 
\begin{cases}
1, \text{ if } E(C) > 2 V + B,\\
0, \text{ if } E(C) < 2 V + B,
\end{cases}
\label{Eq:BGsol}
\end{align}
where $B =\sum_i b_i$. Note that $B$ is a constant of motion (in fact, each $b_i$ is a constant of motion).
If $E(C) = 2 V + B$, $S(C)$ can be either 1 or 0, depending on how energy is distributed. For the present purposes, however, we do not have to care about this case.

Now we choose the initial condition as
\begin{align}
z_i  = b_i + \xi_i,
\label{Eq:sigmai}
\end{align}
where $b_i$ is a random number taking either $1$ or 0 with a restriction $\sum_i b_i = B$ for a certain fixed
$B$ ($0 \le B \le V$) and $\xi_i$'s are independent and identically distributed random 
variables with the common distribution $P_{bg}(n)$ with average $\zeta_0
= \sum_{n=0}^\infty n P_{bg}(n)$.
The mean energy density of the whole system is $\zeta = \zeta_0 + B/V$.
For the moment, we do not specify the form of $P_{bg}(n)$.

As has already been discussed, $w_i$ defined in Eq.~\eqref{Eq:wi} will determine the transition point.
With the above initial condition, the mean value of pairs at site $i$ 
with $b_i = 1$ is 
\begin{align}
\nonumber
2 \langle w_i \rangle &= 
\sum_{k=1}^\infty 2 k \left [ P_{bg}(\xi_i = 2k-1) 
+ P_{bg}(\xi_i = 2k) \right ]
\\
&= \zeta_0 + \sum_{k=0}^\infty P_{bg}(\xi_i = 2k +1) 
= \zeta_0 + p_o,
\end{align}
and at site $j$ with $b_j = 0$ it is
\begin{align}
\nonumber
2 \langle w_j \rangle 
&=\sum_{k=1}^\infty 2 k \left [ P_{bg}(\xi_i = 2k) 
+P_{bg}(\xi_i = 2k+1) \right ] \\
&= \zeta_0 - \sum_{k=0}^\infty P_{bg}(\xi_i = 2k +1) 
=\zeta_0 - p_o,
\end{align}
where $p_o \equiv \sum_{k=0}^{\infty} P_{bg}(2k+1)$ is the probability that $\xi_i$ is an odd number.
Denoting the mean density of pairs by $\langle w \rangle$, we get
\begin{align}
2\langle w \rangle \equiv 2\lambda \langle w_i \rangle + 2(1-\lambda)\langle w_j \rangle
=\zeta_0 - (1 - 2 \lambda) p_o ,
\end{align}
where $\lambda = B/V$. Note that if $P_{bg}(n) \neq 0$ for all $n$,
then the probability for the total number of pairs to exceed $V$ is nonzero
for any finite $\zeta_0$ and, in turn, a phase transition can be well defined
only under the infinite-size limit.

\begin{figure}[t]
\includegraphics[width=\linewidth]{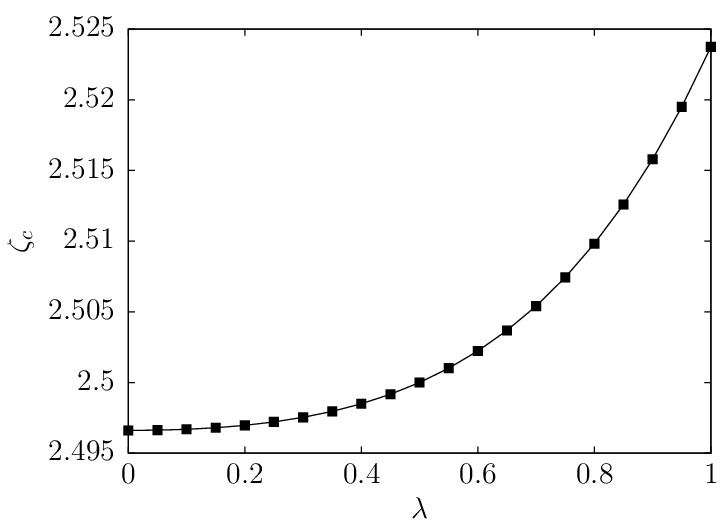}
\caption{\label{Fig:BG} Plot of $\zeta_c$ vs $\lambda$ for the TM~\eqref{Eq:BG} 
with the initial condition \eqref{Eq:sigmai} together
with the distribution \eqref{Eq:sigmai_ini}.}
\end{figure}
Now, we take an infinite $V$ limit with $\lambda $ and $ \zeta_0$ fixed.
If the variance of $P_{bg}$ is finite, the central limit theorem
ensures that the distribution of the density of pairs 
becomes a delta function under this limit. 
Thus, the transition point $\zeta_c$ is determined by the condition 
$\langle w \rangle = 1$, which gives 
\begin{align}
\zeta_c = 2 + \lambda  + (1-2 \lambda) p_o .
\label{Eq:bracelet_sol}
\end{align}
Interestingly, if we choose $\lambda = 0.5$, $\zeta_c$ becomes the SOC critical
density $2.5$~\cite{FLW2010}, regardless of the form of $p_o$.
Except for the case of $\lambda = 1/2$, $\zeta_c$ should depend on $p_o$ and $\lambda$.

To be specific, we solve the case with the Poisson distribution
\begin{align}
P_{bg}(\xi_i = n)  = \frac{\zeta_0^n}{n!} e^{-\zeta_0},
\label{Eq:sigmai_ini}
\end{align}
which gives
\begin{align}
p_o 
= \frac{1-e^{-2\zeta_0}}{2}
= \frac{1-e^{-2(\zeta-\lambda)}}{2}
.
\end{align}
We depict the numerical solution $\zeta_c$ as a function of $\lambda$ in Fig.~\ref{Fig:BG}.
Note that in Ref.~\cite{FLW2010} the above equation with $\lambda = 0$ was studied.

For purposes of later discussions, we also consider another initial condition 
\begin{align}
z_i = d_i + \xi_i,
\end{align}
where $\xi_i$ is the same as in Eq.~\eqref{Eq:sigmai} and $d_i$ is a random number 
that takes either 2 or 3
with the restriction $\sum_i d_i =   D$ ($2V \le D \le 3V$).
Since $\langle w \rangle$ is already $1$ even if $\xi_i = 0$ for all $i$,
$\phi_1  = 1$ for any $\zeta_0 > 0$ in the infinite $V$ limit. 

Notice that a configuration with $z_i = 2$ or $3$ for all $i$ is a recurrent
configuration of the corresponding ASM 
with two dissipative boundary sites at $i=1$ and $i=V$~\cite{FLW2010}.
This example seems to suggest that there is no nontrivial transition 
if an initial configuration of a DFES is constructed by 
adding energy to a recurrent configuration of the corresponding ASM.
Later, we will argue that this is a general feature of the DFES model.

Now, we will discuss a tree. Since
$\zeta_m = \zeta_M$ in the infinite $V$ limit in this case, 
there is a unique transition point $\zeta_c =\zeta_m$ 
if the initial energy density distribution becomes a delta function in the
infinite-size limit. This was observed numerically in Ref.~\cite{FLW2010}
for some trees.

Although we have only used symmetric TMs for illustration,
we would like to emphasize that the existence of many transition points
does not depend on whether or not a TM is symmetric.
\subsection{\label{Sec:phi2}Order parameter $\phi_2$}
To define another order parameter, we consider the steady state density $\rho(C)$ of active sites, 
where $\rho$ is regarded as a function of the initial configuration $C$.
We define $\rho(C)$ for the PU as
\begin{align}
\rho(C) \equiv \lim_{T\rightarrow \infty} \frac{1}{T} \sum_{t=0}^T \frac{n(C_t)}{V},
\end{align}
where $C_t$ is the configuration at (integer) time $t$ with $C_0 = C$ and 
$n(C_t)$ is the number of active sites in the configuration $C_t$.
For the RSU we define $\rho(C)$ as the average density of active sites in the
steady state of the master equation~\eqref{Eq:master} if
the system evolves from the configuration $C$. 
Obviously, $\rho(C) = 0$ irrespective of update rules if $S(C) = 0$.

Although $S(C) = 1$ implies $\rho(C) > 0$,
$\rho(C)$ generally depends on the update rule.
For example, consider a configuration
(0202) for the TM $\tilde F$ in Eq.~\eqref{Eq:BG_red} with $V=4$.
Obviously, $\rho(C) = 1/2$ for the PU. 
In the RSU, there are two possible patterns, (0202) and (1012), up to
translation. Since the average waiting time to the next jump of the first pattern is 1/2 while
that of the second pattern is 1, the probability that the first (second) pattern is found
in the steady state is $1/3$ (2/3). Thus, $\rho(C) =1/3$ when the RSU is employed.

Now, we define the second order parameter $\phi_2$ as
\begin{align}
\phi_2 \equiv \sum_C \rho(C) P_0(C;\zeta,V),
\end{align}
where $P_0$ is the initial distribution and $\zeta$ satisfies Eq.~\eqref{Eq:defzeta}.
As we have shown, $\phi_2$ depends on the update rule,
which should be contrasted with $\phi_1$.
Since $\rho(C) \le S(C)$ for any $C$, $\phi_2$ cannot be larger than $\phi_1$. 
Thus, $\phi_1 = 0$ implies $\phi_2= 0$.
For finite $V$, $\phi_1 \neq 0$ should imply $\phi_2 \neq 0$. 
Thus, a phase transition point for
finite $V$, if it exists, is the same irrespective of whether $\phi_1$ or $\phi_2$ is used as an order parameter.

Although transition points for finite $V$ do not depend on which order parameter 
is used, it is nontrivial to answer whether or not
the infinite-size limit affects this conclusion. 
This is because there are configurations such that 
$\rho(C) \rightarrow 0$ under the infinite $V$ limit, while $S(C) = 1$ for any $V$. 
For example, if
$C = \cdots 1111201111 \cdots$ for the TM \eqref{Eq:BG_red}, $\rho(C) = 1/V \rightarrow 0$ while 
$S(C) = 1$ for any $V$. 
In Appendix~\ref{App:Zm}, we show that it is always possible
to construct such a configuration with $E=Z_m$ for any symmetric TM. 
If such configurations exist for any $E$ in the range $Z_m \le E \le Z_M$, 
we cannot exclude the possibility that 
$\phi_1$ and $\phi_2$ for the same initial condition can give different transition points in
the infinite-size limit. 
Presumably, an ignorant preparation for the initial condition as 
independent and identical Poisson distributions would not generate such
complications. This might be an interesting question, but we will not pursue 
the difference in $\phi_1$ and $\phi_2$ any further in this paper.

\subsection{\label{Sec:phi3} Order parameter $\phi_3$}
In the previous two sections, 
the infinite-size limit, when necessary, is preceded by the infinite-time limit. 
Now we discuss the consequence of changing the order of the two limits.

If the infinite-size limit is taken first, we cannot assign a unique value 
to $S(C)$, introduced in Sec.~\ref{Sec:phi1}, irrespective of the update rule.
For example, consider the TM \eqref{Eq:BG_red} with the following initial configuration
\begin{align}
z_i = 
\begin{cases}
2, & \text{ if } i\pmod{3} =0,\\
0, & \text{ otherwise},
\end{cases}
\end{align}
which resembles $\cdots 0 0 2 0 0 2 0 0 2 0 0 2 \cdots$ in a one-dimensional infinite lattice.
If the PU is employed, the system falls into an absorbing state
in one unit time even if the system size is infinite.
However, if we employ the RSU, that is, the master equation,
the probability that this configuration evolves into an absorbing
state in finite time is zero. 
Still, the density of active sites is zero in the infinite-time limit 
(the average density of active sites at time $t$ in the RSU is $e^{-t}/3$). 

As this example suggests, 
it is appropriate to study the time dependence of the density of active sites.
From the above consideration, we define the third order parameter $\phi_3$ as
\begin{align}
\label{Eq:rhoa}
\rho_a(t) &= \limsup_{V\rightarrow \infty} \sum_C \frac{n(C,V;t)}{V} P_0(C;\zeta,V),\\
\phi_3 &= \lim_{t\rightarrow \infty} \rho_a(t),
\end{align}
where $n(C,V;t)$ is the number of active sites averaged over ensembles
at time $t$ when the system evolves
from the initial configuration $C$.
For convenience, we will exclusively refer to $\rho_a(t)$ as the \emph{activity density} (at time $t$).
Although we did not mention it explicitly, the 
infinite-size limit above should be understood 
as $\limsup$ to guarantee the existence of the limit for any sequence of
$P_0$. 
\begin{figure}[t]
\includegraphics[width=\linewidth]{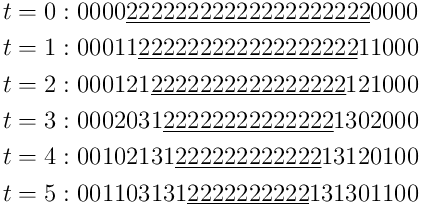}
\caption{\label{Fig:1DEx} Time evolution of the DFES with the TM \eqref{Eq:BG_red}
for the initial configuration \eqref{Eq:1DI} with $V= 80$. 
The PU is employed. At every time step, 
the length of the middle string of 2's (underlined) decreases by 2.}
\end{figure}

It may be tempting to claim that if $S(C) = 0$ for any finite $V$,
the limit of $n(C,V;t)$ should be
\begin{align}
\lim_{t\rightarrow\infty} 
\lim_{V\rightarrow \infty} \frac{n(C,V;t)}{V} = 0.
\end{align}
However, it is not true in general. 
For example, consider again the TM ~\eqref{Eq:BG_red} with an initial configuration $C_0$,
\begin{align}
z_i= 
\begin{cases}
2, &\text{ if } 1 \le i \le V/4, \\
0, &\text{ otherwise.}
\end{cases}
\label{Eq:1DI}
\end{align}
Since the energy density is not larger than $1/2$, the conclusion in Sec.~\ref{Sec:phi1} ensures that $S(C_0) = 0$ for any $V$.

The behavior of $n(C_0,V;t)$ for infinite $V$ can be easily understood by an example; see Fig.~\ref{Fig:1DEx}.
Under the PU, the length of the middle string of consecutive 2's (underlined in Fig.~\ref{Fig:1DEx}) decreases
by 2 at every time step. 
Thus, for any $0<\epsilon \ll 1$, it takes $\epsilon V$  for the string to decrease by $2 \epsilon V$.
Thus, in the infinite $V$ limit, the activity density should remain $1/4$ for any finite
$t$. Hence we get
\begin{align}
\lim_{t\rightarrow\infty} \lim_{V\rightarrow\infty} \frac{n(C,V;t)}{V} = \frac{1}{4},
\end{align}
although $S(C_0) = 0$ for any $V$.

This example shows that, if the infinite $V$ limit is taken before the infinite $t$ limit, 
$\phi_3$ can be positive even if $\zeta < \zeta_m$. 
That is, the transition point need not be 
restricted to be in the range $\zeta_m \le \zeta_c \le \zeta_M$, even 
if Eq.~\eqref{Eq:Edis} is satisfied.
This feature is in fact not limited to the DFES. This conclusion
can be literally applicable to stochastic models; see Sec.~\ref{Sec:dis_sfes}.

If the initial configuration is homogeneous in that any region with volume $\alpha V$
for any $0<\alpha<1$ has the same probability distribution of intensive parameters (such as energy density) as $P_0$ itself in the infinite $V$ limit, 
$\phi_1$ and $\phi_3$ seem to give the same transition point. 
In fact, however, this is not valid unless the TM is also locally the same.
We will clarify this point using an example.

Let us consider a symmetric $2V \times 2V$ TM $G$
constructed by connecting a site $i$ of TM~\eqref{Eq:BG} with $V$ sites to a site $j$
of  TM~\eqref{Eq:BG_red} with $V$ sites such that  $G_{ij} = G_{ji} = -1$,
$G_{ii}=3$, and $G_{jj}=5$ ($1\le i \le V, V < j \le 2 V$). 
In the infinite $V$ limit, $\zeta_m = 3/2$. 
We now consider the initial condition such that the probability that site $i$ has energy $z$ is $\zeta^z e^{-\zeta}/z!$ for
all $i$, where $\zeta=5/4$. It is homogeneous in the sense of the above criterion.
However, $\phi_3$ for this initial condition is nonzero even though 
$\zeta < \zeta_m$, because the activity density in the region governed by the 
TM~\eqref{Eq:BG_red} saturates to a nonzero value if the infinite-size limit
precedes the infinite-time limit.
Hence, a homogeneous initial condition is not sufficient to guarantee 
the existence of a unique transition point irrespective of whether
one studies $\phi_1$ or $\phi_3$.

It does not seem feasible to find a general condition
under which $\phi_1$ and $\phi_3$ can give the same transition point.
Both should be studied separately to fully understand the DFES.
Notwithstanding, space homogeneity in both $P_0$ and the TM seems to give the
same transition point for $\phi_1$ and $\phi_3$, which will be demonstrated
using a concrete model in the next section.

\section{\label{Sec:btwdfes}BTW-FES}
Although the general conclusion in the previous section is in principle applicable
to any DFES, to control the transition point via the initial 
distribution requires explicit information of the set $\A$.
For the bracelet-graph case, this explicit information is available,
which makes it possible to control the transition point at will.
In general, it is difficult, if not impossible, to find the
analytic expression for $\A$. 
Despite the lack of full information about $\A$, however, we will 
show that the existence of various transition points can be easily demonstrated. 

In this section we study the DFES associated with
the two-dimensional BTW model. 
The model, which will be called the BTW-FES, has the following TM
\begin{align}
F_{\bm{x},\bm{y}} = 
\begin{cases}
4, & \bm{x}=\bm{y},\\
-1, & |x_1-y_1|=1 \text{ or } L-1 ,~x_2 = y_2,\\
-1, & |x_2-y_2|=1 \text{ or } L-1 ,~x_1 = y_1,
\end{cases}
\end{align}
where $\bm{x}\equiv (x_1,x_2)$ and $\bm{y}=(y_1,y_2)$ ($x_i ,y_i  = 1,2,\ldots,L$ 
for $i = 1,2$) are two-dimensional vectors for the sites of a square lattice with size $V=L^2$. 

The initial condition we use is
\begin{equation}
z_{\bm{x}} = h_{\bm{x}} + \xi_{\bm{x}}, 
\label{Eq:BTW_ini}
\end{equation}
where $0\le h_{\bm{x}} \le 3$ and $\xi_{\bm{x}}$'s are drawn from  the multinomial distribution 
($M$ is an integer and $\delta$ is the Kronecker delta symbol) 
\begin{align}
\text{Pr}(\{\xi_{\bm{x}}=n_{\bm{x}}\})
= M! \left (\prod_{\bm{x}} \frac{p_{\bm{x}}^{n_{\bm{x}}}}{n_{\bm{x}}!}  \right )
\delta\left (M,\sum_{\bm{x}} n_{\bm{x}}\right ),
\label{Eq:multi}
\end{align}
where $p_{\bm{x}} \ge 0$ for all $\bm{x}$ with the constraint $\sum_{\bm{x}} p_{\bm{x}} = 1$.
In simulations, we set  $h_{\bm{x}}= h_o$ ($h_e$) if $x_1 + x_2$ is odd (even).
That is, the initial configuration is set by 
adding energy to the
checkerboard configuration
\begin{align}
\begin{matrix}
\cdots&\vdots&\vdots&\vdots&\vdots&\cdots\\
\cdots&h_e&h_o&h_e&h_o&\cdots\\
\cdots&h_o&h_e&h_o&h_e&\cdots\\
\cdots&h_e&h_o&h_e&h_o&\cdots\\
\cdots&h_o&h_e&h_o&h_e&\cdots\\
\cdots&\vdots&\vdots&\vdots&\vdots&\cdots
\end{matrix}
\label{Eq:h1h2}
\end{align}
according to the distribution Eq.~\eqref{Eq:multi}.
In the infinite $V$ limit, we keep $\zeta_0 \equiv M/V$ to be constant.
The total energy density is $\zeta = \zeta_0 + (h_o+h_e)/2$, which 
is considered the tuning parameter.

In this section, we will only show simulation results of the case where $p_{\bm{x}} = V^{-1}$ for all $ \bm{x}$, 
which makes the multinomial distribution indistinguishable from
the Poisson distribution,
\begin{align}
\text{Prob}(\{\xi_{\bm{x}}=n_{\bm{x}}\})
= \prod_{\bm{x}} \frac{\zeta_0^{n_{\bm{x}}}}{n_{\bm{x}}!} e^{-\zeta_0},
\end{align}
in the infinite-size limit.
The multinomial distribution with uniform $p_{\bm{x}}$ is implemented by repeating the following procedure
$M$ times: We choose a site at random with equal probability. Then, the energy at that site
increases by one. 

\subsection{\label{Sec:BTWphi1}Phase transition  points: Analysis of $\phi_1$}
\begin{table}[t]
\caption{\label{Table:BTW} Numerical values of $\zeta_{h_o,h_e}$ 
of the BTW-FES for four sets of $h_o$ and $h_e$. The numbers
in parentheses indicate errors of the last digits.
Except for $\zeta_{2,2}$, the estimates for infinite $L$ are from the simulations discussed 
in Sec.~\ref{Sec:macro}. For the discussion as to why $\zeta_{2,2} \rightarrow 2$ as $L\rightarrow \infty$, 
see the text.}
\begin{ruledtabular}
\begin{tabular}{lllll}
$L$&$\zeta_{0,1}$&$\zeta_{1,1}$&$\zeta_{0,2}$&$\zeta_{2,2}$\\
\hline
$2^7$   & 2.124~997(1)&2.133~886(1)&2.117~107(1)&2.133~98(1)\\
$2^8$   & 2.125~023(1)&2.133~902(1)&2.117~133(1)&2.119~88(1)\\
$2^9$   & 2.125~034(1)&2.133~909(1)&2.117~144(1)&2.108~97(2)\\
$2^{10}$& 2.125~038(1)&2.133~912(1)&2.117~147(1)&2.100~21(3)\\
$2^{11}$& 2.125~038(1)&2.133~913(1)&2.117~148(1)&2.092~96(5)\\
$2^{12}$& 2.125~038(2)&2.133~913(1)&2.117~149(1)&2.086~86(9)\\
$\vdots$&$\vdots$&$\vdots$&$\vdots$&$\vdots$\\
$\infty$& 2.125~04(1)&2.133~914(4)  &2.117~15(1)&2
\end{tabular}
\end{ruledtabular}
\end{table}
Employing the method in Ref.~\cite{FLW2010}, we find
the transition point determined by the order parameter $\phi_1$. Note that the case studied in Ref.~\cite{FLW2010} 
corresponds to $h_o = h_e = 0$ in our setting.
To be self-contained, we explain the simulation method~\cite{FLW2010}:
\begin{description}
\item[Step 1] We prepare for the checkerboard configuration Eq.~\eqref{Eq:h1h2}. 
\item[Step 2] We choose a site randomly and increase energy at that site by 1, then
toppling begins.
\item[Step 3] We wait until either the system falls into an absorbing state
or all sites have toppled at least once.
\item[Step 4--1] If the system falls into an absorbing state, we go back to 
step 2 with the present absorbing state.
\item[Step 4--2] If all sites have toppled at least once, the simulation is terminated and we
go back to step 1.
\end{description}
The energy density $e = E/V$ at step 4--2 is a random variable we are interested in and we calculated 
its average, denoted by $\zeta_{h_o,h_e}$, and the standard deviation, denoted by $\sigma_{h_o,h_e}$.

Simulation results for certain combinations of $h_o$ and $h_e$ 
are summarized in Table~\ref{Table:BTW}. 
The error of $\zeta_{h_o,h_e}$ is estimated as $3 \sigma_{h_o,h_e} / \sqrt{N_E}$,
where $N_E$ is the number of realizations for the given parameters.
We also studied the case $h_o=h_e=0$ and obtained a consistent result with Ref.~\cite{FLW2010} (data not shown).
As can be seen from Table~\ref{Table:BTW},
transition points indeed depend on the initial condition.

\begin{figure}[b]
\includegraphics[width=\linewidth]{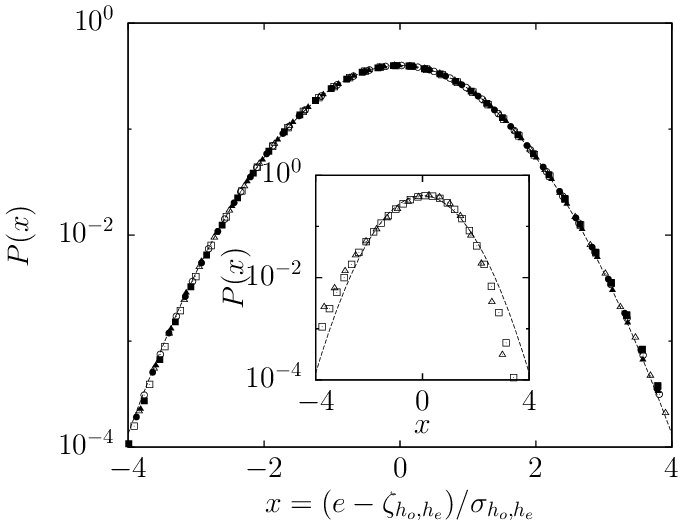}
\caption{\label{Fig:gauss} Semilogarithmic plots of probability density of the reduced parameter
$x = (e - \zeta_{h_o,h_e})/\sigma_{h_o,h_e}$
for $h_o=0$, $h_e=1$ (open symbols) and $h_o=0$, $h_e=2$ (solid symbols).
The systems sizes are $L=2^7$ (squares), $2^8$ (circles), and $2^9$ (triangles). 
For comparison, the Gaussian distribution with zero mean
and unit variance is also drawn by a dashed curve. Inset: Plots of normalized probability density
for $h_o=h_e=2$ on a semilogarithmic scale. The system sizes are $L=2^7$ (squares) and $2^9$ (triangles). The Gaussian distribution which is depicted by a dashed curve  deviates significantly
from $P(x)$. 
}
\end{figure}
The behavior of $\zeta_{2,2}$ is quite distinct from the other cases in two respects.
First, $\zeta_{2,2}$ does not show any symptom of saturation up to $L = 2^{12}$
unlike the other cases, which makes it difficult to estimate the transition point accurately from this type of study. 
Second, as shown in Fig.~\ref{Fig:gauss}, 
the probability distribution of the energy density $e$ is well approximated
by a Gaussian for other cases, while that for $h_o=h_e=2$ deviates significantly
from a Gaussian. 

The configuration Eq.~\eqref{Eq:h1h2} with $h_o=h_e=2$ is 
actually a recurrent configuration of the BTW model.
Now, we will argue that $\phi_1$ in the infinite-size limit 
is nonzero for any nonzero $\zeta_0$
if $h_{\bm{x}}$'s in Eq.~\eqref{Eq:BTW_ini}
form a recurrent configuration of the BTW model.

Whether a stable configuration of the BTW model is recurrent is 
determined by the burning algorithm~\cite{D1990}.
If all sites in a configuration are burnt by the burning algorithm, 
adding energy to every boundary site by the amount of dissipated energy per toppling
there will make all sites topple once and come back to the original configuration.
Therefore, in the context of the BTW-FES, the burning algorithm implies that
the increase of energy density by $4/L$ is enough to make the system active 
if $h_{\bm{x}}$'s form a recurrent configuration.

Since the boundary is a kind of an infinite percolating cluster, 
a random energy distribution with small $\zeta_0$ has no chance
to have such a percolating cluster of sites with added energy,
in the infinite-size limit. 
The burning-algorithm argument can at best provide a crude possibility that 
a negligible density increase might be enough to make the system active.

A more plausible scenario is provided by Ref.~\cite{FLP2010},
where it is shown that the addition of a finite, but large, amount of energy 
(note that the density change is still zero) to the configuration \eqref{Eq:h1h2} with
$h_o=h_e=2$ in the infinite $L$ limit is enough for toppling to continue forever.
To use this observation, let us consider ${\cal N}=N^2$ sites forming a square 
($1\ll N \ll L$). The probability that all these sites are active in the initial 
configuration should be larger than $\zeta_0^{4{\cal N}}$. 
So if $V \zeta_0^{4 {\cal N}} \gg 1$ or
$V \gg \exp[4 {\cal N} \ln(1/\zeta_0)]$, the initial configuration
with high probability has a region with a compact cluster of a large number of 
active sites for nonzero $\zeta_0$. 
Since the toppling dynamics in this region is more or less similar 
to that in Ref.~\cite{FLP2010}, the activity density cannot decay to zero.
That is, the system is in the active phase for any nonzero $\zeta_0$.

\begin{figure}
\includegraphics[width=\linewidth]{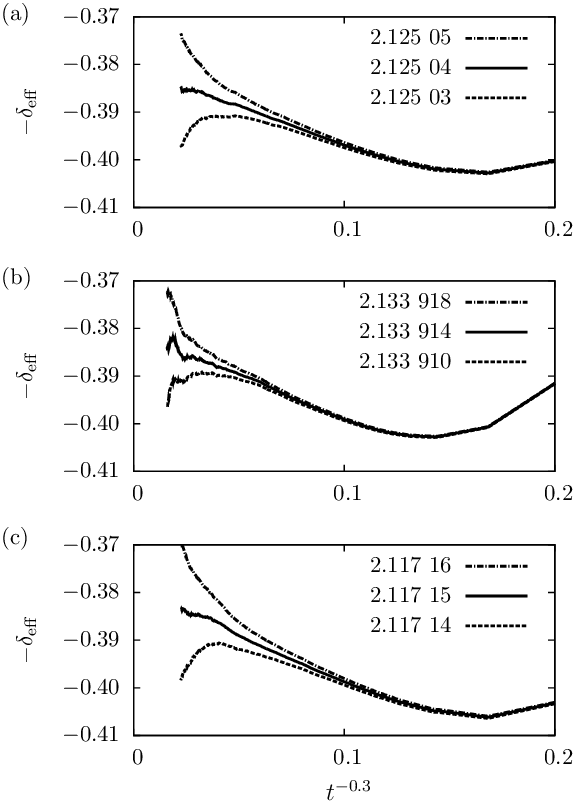}
\caption{\label{Fig:Hall} Plots of the effective exponent
$-\delta_\text{eff}(t)$ around the critical point as
a function of $t^{-0.3}$ for (a) $h_o=0$, $h_e=1$, (b) $h_o=h_e=1$, and (c) $h_o=0$, $h_e=2$.
Although the transition points are different, the decay exponent for every case is $0.38\pm 0.01$,
which supports that all belong to the same universality class.
}
\end{figure}
This argument seems consistent with the numerical results. Assuming
$\zeta_{2,2}$ approaches 2 as $L \rightarrow \infty$,
a naive fitting shows that
\begin{align}
\zeta_{2,2} - 2 \approx 0.65 (\log_2 L)^{-0.8}.
\end{align}
That is, for a system to have nonzero activity for a given $\zeta_0 (\ll 1)$,
the system size $V$ should be much larger than $\exp(\zeta_0^{-1.25})$.
It seems to suggest that the size of the compact cluster
with ${\cal N} \sim \zeta_0^{-1.25}$ would make the system active.
Since the estimate is based on a very crude approximation,
the accuracy should not be taken too seriously.

To confirm, we also simulated the case $h_o=1$, $h_e=3$, which is 
also a recurrent configuration
of the BTW model. As in the case of $h_o=h_e=2$, we observed a 
slow saturation and a non-Gaussian distribution (data not shown). 
Thus, we conclude that a recurrent 
configuration is not a suitable initial condition for studying the APT in the DFES,
once the corresponding ASM has a negligible fraction of boundary sites 
in the infinite-size limit.
\subsection{\label{Sec:macro}Quest for a universality class: Analysis of $\phi_3$}
In this section, we seek an answer for the following three questions. Do $\phi_1$ and $\phi_3$ give the same transition point for the
initial condition Eq.~\eqref{Eq:BTW_ini}? Is the system at the transition point critical? If so, is critical behavior universal? 

These questions can be simultaneously answered by investigating 
the activity density $\rho_a(t)$, which is defined in Eq.~\eqref{Eq:rhoa}. 
If dynamic critical phenomena do exist, the activity density
at the critical point decreases with time in a power-law fashion as
($\delta,\chi >0$)
\begin{align}
\rho_a(t) \sim t^{-\delta} \left [ 1 + \tilde c t^{-\chi} 
+ o(t^{-\chi}) \right ],
\end{align}
where $\delta$ is the critical-decay exponent, $\chi$ is an exponent governing
the leading corrections to scaling, $o(t^{-\chi})$ stands for the collection of all 
terms that decay faster than $t^{-\chi}$, and $\tilde c$ is a constant.
In the active phase where $\phi_3$ is nonzero, $\rho_a(t)$ should eventually
saturate to $\phi_3$. Meanwhile, in the absorbing phase where $\phi_3$ is zero,
$\rho_a(t)$ decays to zero exponentially (or at least faster than
the critical decay $t^{-\delta}$).

The different behavior in each phase is clearly displayed by
the effective exponent 
\begin{equation}
-\delta_\text{eff}(t) \equiv \frac{\ln[\rho_a(t)/\rho_a(t/b)]}{\ln b},
\label{Eq:Def_eff}
\end{equation}
where $b$ is a constant larger than 1.
At the critical point, the effective exponent in the asymptotic regime
should behave as
\begin{align}
-\delta_\text{eff}(t) = -\delta - c t^{-\chi} + o(t^{-\chi}),
\label{Eq:deff_cri}
\end{align}
where $c= \tilde c (b^\chi - 1)/\ln b$.
If we plot $-\delta_\text{eff}(t)$ as a function of $t^{-\chi}$ with 
the correct value of $\chi$,
the effective exponent at the critical point becomes a straight line
for large $t$, which should cross the ordinate at $-\delta$. 
On the other hand, if the system is in the active (absorbing) phase 
the effective exponent should eventually veer up (down) 
as $t^{-\chi}$ decreases.

Now, we present the simulation results. 
The system size in our simulations is $60~000\times 60~000$
(that is, $L=60~000$) and we observed the dynamics up to
$t_\text{max} = 10^6$ for the case of $h_o=h_e=1$ and up to $
t_\text{max}=3 \times 10^5$  for the cases of $h_o=0$, $h_e=1$ and
$h_o=0$, $h_e=2$. 
To ensure that we indeed observed the infinite-size limit,
we also simulated the system with smaller size ($L=20~000$, actually),
to find almost identical behavior up to the observation time (data not shown).
We also checked whether the asymptotic behavior can depend on the update rule.
We found that the statistically significant difference is observed only 
for short time ($t\le 10$) and the long time behavior is almost identical.
So we only show the behavior of the system evolving according to the PU.

Figure~\ref{Fig:Hall} shows the behavior of the effective exponents 
with $b=10$ for (a) $h_o=0$, $h_e=1$, (b) $h_o=h_e=1$, and 
(c) $h_o=0$, $h_e=2$.
The activity density for each parameter set is the result of the average over 100 
independent runs. 
Although Fig.~\ref{Fig:Hall} uses $\chi=0.3$, varying $\chi$ from 0.2 to 0.4
also makes the middle curve resemble a straight line.
Estimating the error of the exponents by the range of extrapolated values 
for different $\chi$, we concluded that
the critical behavior is indeed universal with $\delta = 0.38\pm 0.01$.
The estimated critical energy density for each case is listed
on the last line of Table~\ref{Table:BTW},
which strongly supports that $\phi_1$ and $\phi_3$ give the same critical point.
Hence, the answers to the three questions raised in the beginning of this section
are all positive.
Notice that in Ref.~\cite{VDMZ2000} $\delta$ was estimated as 
0.41 for the case of $h_o=h_e=0$. 
Since the critical point was set to be $2.125$ in Ref.~\cite{VDMZ2000}, 
which actually corresponds to the absorbing phase~\cite{FLW2010}, 
slightly larger value than ours is consistent with our estimate. 

\begin{figure}[t]
\includegraphics[width=\linewidth]{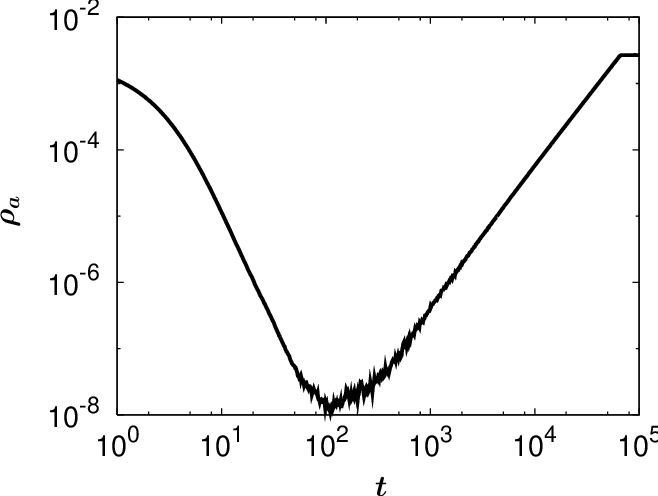}
\caption{\label{Fig:H2a} Plot of $\rho_a$ vs $t$ for $h_o=h_e=2$ and 
$\zeta = 2.07$. 
}
\end{figure}

When $h_o=h_e=2$, we claimed that the system is in the active phase
for any $\zeta>2$. Now we discuss the behavior of the activity density in this 
case. When $\zeta$ is close to 2, the initial activity 
density is very small [approximately $(\zeta-2)^2/2$] and the typical
distance between two active sites in the initial configuration is 
$\sim 1/(\zeta-2)$. So the activity density is likely to decrease very 
quickly in the early stage. 
But eventually the initial compact cluster of active sites, discussed
at the end of the previous subsection, will play a
dominant role and the activity should spread throughout the system.
To confirm this anticipation,
we simulated the system with $L = 20~000$ for $h_o=h_e = 2$.
The result of eight independent runs is shown in Fig.~\ref{Fig:H2a}.  
As anticipated, the density decays very quickly up to $\rho_a\approx 10^{-8}$.
Actually, six out of eight runs lost all active sites around $t=100$. 
However, the activity density in the two remaining runs eventually increased and 
saturated to a finite value, as anticipated.
As this example shows, it is very hard to confirm numerically that the system 
is in an active phase for very small $\zeta_0$.

\subsection{\label{Sec:mapt}Microscopic absorbing phase transitions}
In this section, we study a phase transition of `avalanche' dynamics 
initiated by adding a unit energy to a randomly chosen site in an 
absorbing state of an infinite system. Since the system is infinite, 
the location of the site chosen for adding energy in the beginning 
will be set as the origin without loss of generality. We are now 
interested in whether the avalanche
dynamics can continue forever or will be terminated in finite time.
Unlike the previous study, the (macroscopic) \emph{density} of 
active sites is always zero.  In this regard, we will call this kind of a 
phase transition a microscopic absorbing phase transition (\miAPT), first 
coined in Ref.~\cite{PP2008EPJB}.  For comparison, a phase transition 
in the previous sections, which occurs with nonzero activity density in the infinite-size 
limit, will be called a macroscopic APT (\maAPT).

The infinite-time
limit of the survival probability $P_s(t)$ plays the role of the order 
parameter. $P_s(t)$ is defined as the probability that the system
has an active site at time $t$.
We also study the mean number of active sites $N(t)$ at time $t$ 
that is averaged over \emph{all} ensembles and the mean spreading $R^2(t)$ of 
the \emph{surviving} ensemble. In simulations, we measure the mean spreading as
\begin{align}
R^2(t) = \left \langle \left ( x_\text{max} - x_\text{min} \right )^2 \right \rangle_s,
\end{align}
where $x_\text{min} (x_\text{max})$ is 
the  minimum (maximum) among the first components of 
active-site vectors at time $t$ and $\langle \cdots \rangle_s$ stands for the average 
over the surviving ensemble.
If the system is critical at the transition point, these quantities 
have asymptotic power-law behaviors as
\begin{align}
P_s(t) \sim t^{-\theta},\quad
N(t) \sim t^\eta,\quad
R^2(t) \sim t^{z},
\label{Eq:sp_exp}
\end{align}
where $\theta$, $\eta$, and $z$ are the spreading exponents.

For a concrete discussion, we now specify how
the initial absorbing state is constructed.
For each site $\bm{x} \in Z^2$, $z_{\bm{x}}$ is assigned 0, 1, 2, or 3 
with probability $(1-p_3)p_0$, $(1-p_3)p_1$, $(1-p_3)p_2$, and 
$p_3$, respectively, where $0 \le p_i \le 1$ and $p_0 + p_1 + p_2 = 1$. 
The corresponding energy density of the system is $\zeta = (1-p_3)(p_1 + 2 p_2) + 3 p_3$, which is in the range $0\le \zeta \le 3$.
After constructing the initial absorbing state,
the energy at the origin $\bm{0}$ increases by one ($z_{\bm{0}} \mapsto 
z_{\bm{0}}+1$).
If the origin becomes active by this addition, 
which happens with probability $p_3$, 
the avalanche dynamics begin. 

If $p_3$ is larger than the site percolation threshold $p^* = 0.592~746$~(see,
for example, Ref.~\cite{Stauffer1992}), the initial absorbing state contains 
an infinite percolating cluster of 3's. 
In this section, a cluster will mean a connected cluster
of sites with energy $3$.
If the origin happens to be in the 
infinite cluster, toppling can persist forever at least within this cluster. 
Hence, $P_s(t)$ for
$p_3 > p^*$ should saturate to a nonzero value that cannot be smaller than 
the probability that the origin is a member of the infinite percolating
cluster. Thus, the energy density $\zeta_c$ at the transition point 
is always bounded by
\begin{equation}
\zeta_{c} \le \zeta_\text{bound} \equiv 
(1-p^*) ( p_1 + 2 p_2) + 3 p^*.
\end{equation} 
Note that $\zeta_\text{bound}$ ranges
from $3 p^* \approx 1.78$ (when $p_0=1$) to $2 + p^* \approx 2.59$
(when $p_2=1$).
It is remarkable that the energy density 
at a transition point can be smaller than $\zeta_m = 2$
even though the initial distribution is homogeneous in space. Since the energy 
density of any recurrent configuration of the two-dimensional BTW model cannot be smaller than 
2, the critical avalanche dynamics of the \miAPT with $\zeta_c$ 
smaller than 2 has nothing to do with the SOC behavior.

A clear connection of the \miAPT to the dynamical isotropic
percolation can be seen if we set $p_0=1$ ($p_1 = p_2=0$) and use
$p_3$ as a tuning parameter. 
In this case, any finite cluster is surrounded by (perimeter) sites with zero energy.
If the origin happens to be a member of a finite cluster,
the avalanche dynamics should be terminated within a finite time,
because toppling at the boundary sites of sparse clusters 
cannot make any perimeter site active.
Hence, $P_s(t)$ in the infinite-time limit is nonzero if and only if
an infinite percolating cluster appears with nonzero probability. Hence, we
conclude $p_c=p^*=0.592~746$ for $p_0=1$.

At the percolation threshold, a large compact cluster hardly 
appears~\cite{Stauffer1992}.  Hence, toppling more than once at a given site is 
very unlikely. In this sense, the avalanche dynamics is almost identical to the 
general epidemic process (GEP)~\cite{ABS:M1977,ABS:G1983}
or the dynamical isotropic percolation process, which is characterized by the critical 
exponents~\cite{MDVZ1999}
\begin{align}
\theta = 0.092,\quad \eta = 0.586,\quad z = 1.771.
\label{Eq:maptex}
\end{align}

If $p_0<1$, the existence of a percolating cluster is not necessary
 for the system to be in an active phase.
This can be easily understood by an example.
Consider the following configuration,
\begin{align}
\begin{matrix}
\cdots&\vdots&\vdots&\vdots&\vdots&\vdots&\cdots\\
\cdots&0&0&3&0&0&\cdots\\
\cdots&0&\underline{3}&h&\underline{3}&0&\cdots\\
\cdots&0&\underline{3}&\underline{4}&\underline{3}&0&\cdots\\
\cdots&0&0&0&0&0&\cdots\\
\cdots&\vdots&\vdots&\vdots&\vdots&\vdots&\cdots
\end{matrix}
\label{Eq:ex_mapt}
\end{align}
where $h$ is smaller than 3. If $h=0$, toppling will end after all 
underlined sites topple once.
On the other hand, if $h>0$, this site will be active, which again makes
a site in another cluster (in this example, the site
with energy 3 in the first row) active. That is, if $p_0$ is
smaller than one, the avalanche dynamics starting in a finite cluster can 
trigger an avalanche in another cluster. Accordingly,
the critical point $p_c$ should be smaller than $p^*$.

\begin{figure}[t]
\includegraphics[width=\linewidth]{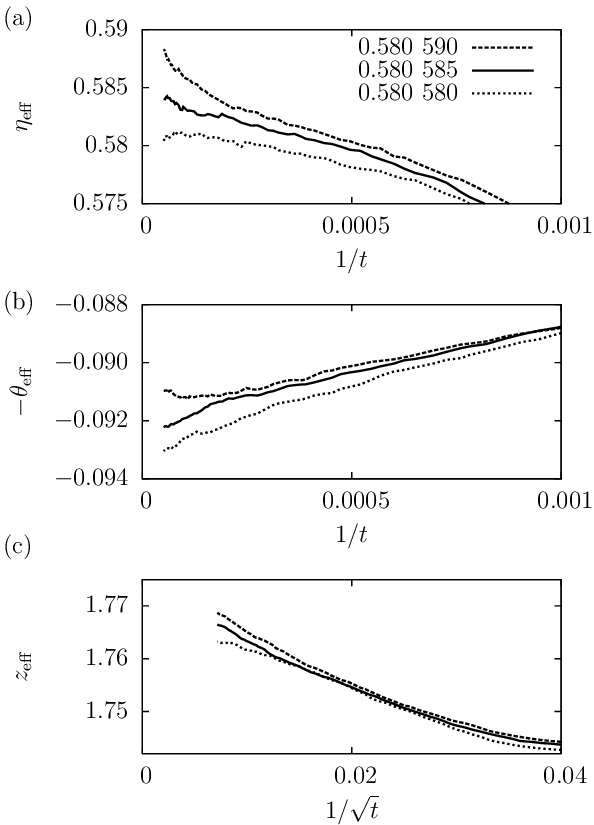} 
\caption{\label{Fig:mapt} Plots of (a) $\eta_\text{eff}$ vs $1/t$, (b) $-\theta_\text{eff}$ vs $1/t$, and (c) $z_\text{eff}$ vs $1/\sqrt{t}$ for
the avalanche dynamics with  Eq.~\eqref{Eq:socden}. 
The values of $p_3$ are 0.580~59 (top curve), 0.580~585 (middle curve), and 0.580~58 (bottom curve).
The critical exponents are consistent with those of the dynamical
isotropic percolation, given in Eq.~\eqref{Eq:maptex}.}
\end{figure}
Although $p_c$ is strictly smaller than the percolation threshold,
the avalanche dynamics at $p_c$ is likely to be the same as the GEP. 
To check this anticipation, we numerically studied the \miAPT for $p_0<1$.
Since $p_3$ rather than the total density plays an important role, we fix
the total density as 17/8 and
use $p_3$ as a tuning parameter.  To be specific, we set $p_2 = 0$ and
\begin{align}
p_1 = \frac{17-24p_3}{8(1-p_3)},\quad
p_0 = 1-p_1 - p_2,
\label{Eq:socden}
\end{align} 
with $0.5625=9/16 \le p_3 \le 17/24\approx 0.71$.
In simulations, we only considered the cases in which the energy at the origin is 
4 in the beginning. Using the RSU, we simulated $6\times 10^6$ independent runs 
for each 
parameter up to the maximum observation time $2\times 10^4$.
For the analysis, we calculated the effective exponents, $-\theta_\text{eff}$,
$\eta_\text{eff}$, and $z_\text{eff}$, defined similarly to Eq.~\eqref{Eq:Def_eff}
with $b=4$.
In Fig.~\ref{Fig:mapt}, we plot the effective exponents (a) $\eta_\text{eff}$, 
(b) $-\theta_\text{eff}$, and (c) $z_\text{eff}$. 
From this figure, we conclude that the critical point is $p_c = 0.580~585(5)$
and the critical exponents are indeed consistent with Eq.~\eqref{Eq:maptex}.

\section{\label{Sec:Dis} Discussion}
\subsection{\label{Sec:hyperscaling}Failure of the hyperscaling relation}
In many APT models, the \miAPT is  deeply related to the \maAPT, 
which is manifest by the hyperscaling relation~\cite{GT1979}
\begin{align}
\eta + \delta + \theta = dz/2,
\label{Eq:hyper}
\end{align}
where $d$ is the dimension. Plugging the \miAPT exponents in Eq.~\eqref{Eq:maptex} and the \maAPT exponent $\delta\approx 0.38$ into Eq.~\eqref{Eq:hyper} with $d=2$, 
one can see that the hyperscaling relation is not valid for the BTW-FES.

The failure of the hyperscaling relation can be interpreted as a lack
of a connection between the \miAPT and \maAPT for the BTW-FES.
In the \maAPT, the active phase in terms of $\phi_1$
is characterized by a compact cluster of toppled sites~\cite{FLW2010}, 
whereas in the \miAPT the existence of a percolating
cluster of toppled sites, which is usually sparse, is sufficient for the 
system to be in the active phase. Since
the (fractal) structures of the toppled sites are different, 
the hyperscaling relation cannot be valid in the BTW-FES model.

The initial absorbing state of the \miAPT in Sec.~\ref{Sec:mapt}
is far from the recurrent configurations of the SOC model, 
whereas in the \maAPT the system at the critical point
most likely approaches one of the recurrent configurations.
In this context, it is still to be answered whether the hyperscaling
relation may be restored if the initial absorbing state in the \miAPT is
a recurrent configuration. The recurrent configurations, however, do not seem
to provide a nontrivial \miAPT, as in Sec.~\ref{Sec:BTWphi1}.

To clarify, let us consider the case with $p_2=1$. Since all sites
have energy either 2 or 3 for any $p_3$, the initial absorbing state
is a recurrent configuration. For any finite $p_3$, the probability that the origin
is a member of a compact cluster of 3's with size ${\cal N}$ is about $p_3^{\cal N}$,
which is nonzero. Once the origin is located in this compact cluster, 
the avalanche dynamics would be
very similar to that studied in Ref.~\cite{FLP2010}, which shows that
finitely many grains are enough to continue toppling forever when
all background sites have energy 2. Accordingly, we can conclude that
for any finite $p_3$, the survival probability $P_s(t)$ should saturate to a 
nonzero value and there is no nontrivial transition. Our preliminary simulation is 
indeed consistent with this conclusion (data not shown). 
This should be paralled with the absence of a nontrivial phase transition 
for $h_o=h_e=2$ in Sec.~\ref{Sec:BTWphi1}. Thus, we do not expect that the
hyperscaling can be restored by another choice of an initial condition.

From this consideration, we can conclude that the \miAPT, which looks similar
to the avalanche dynamics of the corresponding ASM right after the external drive,
has nothing to do with SOC.
Indeed, we have seen that \miAPT of the BTW-FES model is associated
with the dynamical isotropic percolation universality class, 
or the GEP, rather than the SOC universality class.
\subsection{Deterministic Abelian sandpile models can have various critical densities}
In the conventional ASM, all recurrent configurations are equally probable to 
occur in the SOC state
as far as only a single grain is added per drive 
and the probability of any site to be chosen for a drive is nonzero~\cite{D1990} 
(in this section, we are using the terminology of the ASM in such a way 
that a single grain here corresponds to the unit energy in the DFES and 
height at a certain site corresponds to the energy at that site, and so on).
So it might sound plausible to conclude that it
is the special property of the DFES to have various transition points.

But, what happens if the addition of a grain in the drive is correlated with 
a (stable) configuration? As a concrete example, we introduce the possibility 
that a grain added in the drive 
is reflected off many times until it finds a place to be added.
To be specific, we consider the following reflection scheme
in the two-dimensional BTW model (a generalization to an
arbitrary ASM is straightforward)~\cite{DharPri}.
If a grain is to be added in the drive to a 
site, say, $\bm{x}$, with a height of less than 3, the addition is successful with 
certainty.  On the other hand, if the energy at site $\bm{x}$ is 3, the 
grain is added there with probability $p$. With probability $1-p$, however, this 
grain is reflected off to another site chosen at random with equal probability
among all sites and then repeats the above attempt.

The case with $p=1$ corresponds to the conventional setting of the BTW model.
If $p=0$, the drive always ends up with the configuration 
with $z_i = 3$ for all $i$.  If $p>0$, 
all recurrent configurations of the conventional BTW model 
are again recurrent, but the measure is not likely to 
be uniform. Hence, the critical grain density should depend on $p$, although the 
toppling rule in the avalanche dynamics is the same. 
To conclude, the ASM in general and the BTW in 
particular can have various critical densities, by
changing the invariant measure as illustrated above.

Since dissipation on the boundary sites is still inevitable for any nonzero $p$, 
it is very likely that the avalanche can be huge and the 
probability distribution of avalanche size would have a power-law behavior. 
Since the criticality of this model is beyond the scope of this 
paper, we would like to defer it to a later publication.

\subsection{\label{Sec:dis_sfes}Stochastic fixed-energy sandpile models}
Since the existence of $\mI$ and $\A$ with an overlapped energy (density)
range is the origin of the various transition points in the DFES,
a stochastic model with this property should have many transition 
points. Such a stochastic model can be easily constructed 
by employing the idea of the bracelet graph. To illustrate,
let us consider the fixed-energy variant of the 
 one-dimensional (Abelian) Manna model~\cite{M1991}, which we will
call the Abelian Manna fixed-energy sandpile model (AM-FES).
In the AM-FES, a site with energy $z_i$ higher than 1 is active.
An active site topples and  stochastically distributes energy to 
its nearest neighbors, which changes the configuration such that
\begin{align}
&z_i \mapsto z_i -2,\nonumber \\
&\begin{cases}
z_{i+1} \mapsto z_{i+1}+2,&\text{ with probability } 0.25, \\ 
z_{i-1} \mapsto z_{i-1}+2,&\text{ with probability } 0.25,\\ 
z_{i\pm 1} \mapsto z_{i\pm 1}+1, &\text{ with probability } 0.5,
\end{cases}
\end{align}
where periodic boundary conditions are assumed ($1 \le i \le V$).

If the total energy is lower than $V$, the system eventually falls into an 
absorbing state, because the probability current from any configuration to an 
absorbing state is nonzero (the system can behave with small but
nonzero probability as the deterministic model until it falls into an 
absorbing state). On the other hand, if the total energy
is higher than $V$, the system evolves forever, because the system does not have
an absorbing state. 
So, if we use $\phi_1$ as defined in Sec.~\ref{Sec:phi1}, the transition
point is $\zeta_c =1$. If we use $\phi_3$ as defined in
Sec.~\ref{Sec:phi3}, however, it is
known that the transition point is $\zeta_c < 1$~\cite{DAMPVZ2001}.
Here, we assume that the initial energy at every site
is an independent and identically Poisson-distributed random number
with mean $\zeta$. 

Now we consider the bracelet-graph-type modification
of the AM-FES. In this version, a site is active if its energy is 4 or 
higher. When an active site $i$ topples, the energy at site $i$ will be
distributed as
$z_i \mapsto z_i - 4 $ and
\begin{align}
z_{i+1} \mapsto z_{i+1}+4,&\text{ with probability } 0.25,\nonumber \\ 
z_{i-1} \mapsto z_{i-1}+4,&\text{ with probability } 0.25,\\ 
z_{i\pm 1} \mapsto z_{i\pm 1}+2,
&\text{ with probability } 0.5. \nonumber
\end{align} 
Just as the bracelet graph, $b_i \equiv z_i \pmod{2}$ for every
$i$ is a constant of motion. 
If we use $\phi_1$ as the order parameter, 
the transition point is the same as the DFES version of the bracelet
graph. That is, the transition point should depend on the initial condition.
Even if we use $\phi_3$ as an order parameter, 
the transition point depends on the initial condition, too. 
Hence, the SFES can in principle have a different critical density from
the SOC density.

In the AM-FES, spatial inhomogeneity in the initial condition 
can be introduced as in Eq.~\eqref{Eq:multi} by a different
choice of $p_{\bm{x}}$, where $\bm{x}$ should be understood
as a one-dimensional vector. For example,
we can choose $p_{\bm{x}} = 1/M$ if $|\bm{x}| \le M$ and $p_{\bm{x}} = 0$ otherwise, where
$M/V \rightarrow \alpha < 1$ as $V \rightarrow \infty$. 
Since the volume of sites with nonzero energy increase at best linearly in
time in a similar way to Fig.~\ref{Fig:1DEx}, 
the high density region is almost decoupled from the low density region
in the infinite-size limit and the system
can be regarded as consisting of two independent subsystems.
Since the density is calculated in the whole system while
the activity remains in one subsystem, the transition
point can be adjusted at will by the size of two subsystems.
\section{\label{Sec:sum}Summary}
Up to now, we have shown that deterministic fixed-energy sandpile models
can have various transition points, depending on the initial condition as well as on 
the definition of the order parameter.
As a nontrivial example, we numerically studied the Bak-Tang-Wiesenfeld-type 
fixed-energy sandpile model in two dimensions. 
For the initial condition of the BTW-FES, we add energy in a multinomially
distributed way to a stable configuration. By changing
the stable configuration to which energy is added for constructing
an initial configuration, we have observed various transition
points. Furthermore, we have observed and argued that, if
the prepared initial configuration is a recurrent configuration of the conventional
two-dimensional BTW model, any finite density increase makes the system
active and there is no nontrivial phase transition.  

Although the BTW-FES has various transition points, we showed numerically that
there are critical phenomena that are universal irrespective
of the transition point. We also studied the so-called microscopic
absorbing phase transition, which looks similar to the avalanche dynamics
of the BTW model after a drive.
We showed that the critical behavior in the \miAPT is also universal but 
the universality class is the dynamical isotropic percolation class, 
rather than the SOC universality class. As in the study of the \maAPT, 
we argued that there is no nontrivial \miAPT if the initial configuration before 
adding a unit energy is a recurrent configuration.

\begin{acknowledgments}
The author thanks D. Dhar, A. Fey, and J. Krug for helpful discussions.
This work was supported by the Basic Science Research Program through the
National Research Foundation of Korea~(NRF) funded by the Ministry of
Science and ICT~(Grant No. 2017R1D1A1B03034878).
The author thanks the Regional Computing Center of
the University of Cologne (RRZK) for providing computing time on the 
DFG-funded High Performance Computing system CHEOPS as well as for support.
He is also grateful for the hospitality of the Max-Planck-Institut f\"ur Physik komplexer Systeme in Dresden, Germany, where part of this work was completed.
\end{acknowledgments}

\appendix
\section{\label{App:Zm}$Z_m$ FOR A SYMMETRIC TM}
For a symmetric TM $F$, consider the configuration $C_m$ with
$z_1 = F_{11}$ and
\begin{align}
z_i = F_{ii} + \sum_{k=1}^{i-1} F_{ki},
\end{align}
for $i\ge 2$.
Since site $1$ is active, this site is going to topple.
Once sites from 1 to $i-1$ have toppled, 
$z_{i}$ becomes $F_{ii}$ and it will topple.
Hence, every site will topple one after another, which gives $S(C_m)=1$ and
$\rho(C_m) = 1/V$.

Now we will show that the total energy of the configuration $C_m$ is 
$Z_m$ for a symmetric TM. That is,
\begin{align}
Z_m = \sum_i F_{ii} + \sum_{i=2}^V \sum_{k=1}^{i-1} F_{ki}
=\frac{1}{2} \sum_i F_{ii}.
\end{align}
If $S(C)=1$,
there should be a sequence of toppling 
along which all sites topple exactly once and come back to the 
initial configuration.
Since the amount of energy
moving between $i$ and $j$ during the above toppling sequence 
is $|F_{ij}| = |F_{ji}|$, the total energy cannot be smaller than 
\begin{align}
Z_m = \sum_{i,j,i\neq j} \frac{1}{2}| F_{ij}|  = \frac{1}{2} \sum_i F_{ii},
\label{Eq:Zmdef}
\end{align}
where the first summation is over all $i$ and $j$ with $i\neq j$
and we have used $F_{ii} = \sum_{j\neq i} |F_{ij}|$.
Since we have already shown $C_m \in \A$, the proof is complete.
\section{\label{App:tree}PROOF OF $Z_m=Z_M+1$ FOR A TREE}
In this Appendix, we are interested in $Z_m$ for a symmetric TM 
satisfying the following two conditions:
\begin{description}
\item[Condition 1] $|F_{ij}|\le 1$ for all $i\neq j$. 
\item[Condition 2] For any ordered pair of two different site indices $(i,j)$ $(1\le i,j\le V)$, there is a {\em unique} ordered set of different indices $(k_1,k_2,\ldots,k_{n-1})$
such that 
$\prod_{l=0}^{n-1} F_{k_l k_{l+1}}  \neq 0$,
where $k_0 \equiv i$ and $k_n \equiv j$. 
\end{description}
\textbf{Condition 2} is almost the same as the definition of
an irreducible TM, except that the set of indices is unique.
We will refer to a TM satisfying the above two conditions as a tree and
we will prove that $Z_m = Z_M+1$ for any tree.

For the proof, we use the induction.
Assume that $Z_m = Z_M+1$ for a $V\times V$ tree $F$.
Now, choose an arbitrary $i$ ($1\le i \le V$) 
and construct a $(V+1)\times
(V+1)$ TM $\tilde F$ in such a way that 
$\tilde F_{ii} = F_{ii}+1$, $\tilde F_{V+1,V+1} = 1$,
$\tilde F_{i,V+1}=\tilde F_{V+1,i}=-1$, 
$\tilde F_{j,V+1}=\tilde F_{V+1,j} = 0$ ($j \neq i$),
and $\tilde F_{kl} =F_{kl}$ if
neither $k$ nor $l$ is $V+1$ except $k=l=i$.
Obviously, $\tilde F$ is a tree.

Since $\tilde F_{ii}$ increases by one and $\tilde F_{V+1,V+1}=1$, 
$Z_M$ increases by 1.
Also, by the proof in Appendix~\ref{App:Zm}, $Z_m$ also increases by 1.
Thus the relation $Z_m = Z_M+1$ also holds for $\tilde F$.
Since $Z_m = 1$ and $Z_M = 0$ for a tree
\begin{align}
F_2 \equiv \begin{pmatrix}
1&-1\\-1&1
\end{pmatrix},
\label{Eq:tree2}
\end{align}
any tree constructed from $F_2$ by adding sites as in
the induction step should satisfy $Z_m = Z_M+1$.

For a given $V\times V$ tree, we can construct a $(V-1)\times (V-1)$ 
tree by eliminating a site with $F_{ii}=1$. If we continue
the elimination until only two sites remain, this procedure will end up with 
$F_2$. Thus, by tracing back the elimination procedure to
the original TM, we can conclude that $Z_m = Z_M+1$ for any tree.

Now we will show that the relation $Z_m = Z_M+1$ is valid only for a
tree. If a symmetric TM $F$ is not a tree, there are two sites $i$, $j$ such that
two different ordered sets of 
different indices $(k_1,\ldots,k_{n-1})$ and $(l_1,\ldots,l_{m-1})$ with 
\begin{align}
\prod_{p=0}^n F_{k_{p}k_{p+1}} \neq 0, \quad
\prod_{q=0}^m F_{l_{q}l_{q+1}} \neq 0,
\end{align}
where $k_0=l_0= i$ and $k_n = l_m = j$, exist. We set $n\ge m$ without losing 
generality. If there is $p$ such that
$l_p \neq k_p$ and $l_i = k_i$ for all $i$ with $0<i<p\le m$,
we set $x = l_p$ and $y = l_{p-1}$.
If such $p$ does not exist, we set $x = l_m$ and $y = l_{m-1}$.
Now we construct a symmetric and irreducible TM $G$ such that
$G_{xx}=F_{xx}-1$, $G_{yy} = F_{yy}-1$,
$G_{xy} = G_{yx} = F_{xy}+1$, and
$G_{ij} = F_{ij}$ if either $i$ or $j$ is different from $x$ and $y$.
Since $Z_M(G) = Z_M(F)-2$ and  $Z_m(G) = Z_m(F)-1$,
$Z_m(G)$ becomes larger than $Z_M(G)+1$ if $Z_m(F) = Z_M(F)+1$.
Since this is not possible, $Z_m= Z_M+1$ cannot be valid if a (symmetric) TM is not 
a tree.

\bibliography{../../Biblio/prebib.bib,../../Biblio/abs.bib,../../Biblio/soc.bib,../../Biblio/book.bib}

\begin{thebibliography}{26}%
\makeatletter
\providecommand \@ifxundefined [1]{%
 \@ifx{#1\undefined}
}%
\providecommand \@ifnum [1]{%
 \ifnum #1\expandafter \@firstoftwo
 \else \expandafter \@secondoftwo
 \fi
}%
\providecommand \@ifx [1]{%
 \ifx #1\expandafter \@firstoftwo
 \else \expandafter \@secondoftwo
 \fi
}%
\providecommand \natexlab [1]{#1}%
\providecommand \enquote  [1]{``#1''}%
\providecommand \bibnamefont  [1]{#1}%
\providecommand \bibfnamefont [1]{#1}%
\providecommand \citenamefont [1]{#1}%
\providecommand \href@noop [0]{\@secondoftwo}%
\providecommand \href [0]{\begingroup \@sanitize@url \@href}%
\providecommand \@href[1]{\@@startlink{#1}\@@href}%
\providecommand \@@href[1]{\endgroup#1\@@endlink}%
\providecommand \@sanitize@url [0]{\catcode `\\12\catcode `\$12\catcode
  `\&12\catcode `\#12\catcode `\^12\catcode `\_12\catcode `\%12\relax}%
\providecommand \@@startlink[1]{}%
\providecommand \@@endlink[0]{}%
\providecommand \url  [0]{\begingroup\@sanitize@url \@url }%
\providecommand \@url [1]{\endgroup\@href {#1}{\urlprefix }}%
\providecommand \urlprefix  [0]{URL }%
\providecommand \Eprint [0]{\href }%
\providecommand \doibase [0]{http://dx.doi.org/}%
\providecommand \selectlanguage [0]{\@gobble}%
\providecommand \bibinfo  [0]{\@secondoftwo}%
\providecommand \bibfield  [0]{\@secondoftwo}%
\providecommand \translation [1]{[#1]}%
\providecommand \BibitemOpen [0]{}%
\providecommand \bibitemStop [0]{}%
\providecommand \bibitemNoStop [0]{.\EOS\space}%
\providecommand \EOS [0]{\spacefactor3000\relax}%
\providecommand \BibitemShut  [1]{\csname bibitem#1\endcsname}%
\let\auto@bib@innerbib\@empty
\bibitem [{\citenamefont {Tang}\ and\ \citenamefont {Bak}(1988)}]{TB1988}%
  \BibitemOpen
  \bibfield  {author} {\bibinfo {author} {\bibfnamefont {C.}\ \bibnamefont
  {Tang}}\ and\ \bibinfo {author} {\bibfnamefont {P.}\ \bibnamefont {Bak}},\
  }\bibfield  {title} {\bibinfo {title} {Critical Exponents and
  Scaling Relations for Self-Organized Critical Phenomena}, }\href {\doibase
  10.1103/PhysRevLett.60.2347} {\bibfield  {journal} {\bibinfo  {journal}
  {Phys. Rev. Lett.}\ }\textbf {\bibinfo {volume} {60}},\ \bibinfo {pages}
  {2347} (\bibinfo {year} {1988})}\BibitemShut {NoStop}%
\bibitem [{\citenamefont {Dickman}\ \emph {et~al.}(1998)\citenamefont
  {Dickman}, \citenamefont {Vespignani},\ and\ \citenamefont
  {Zapperi}}]{DVZ1998}%
  \BibitemOpen
  \bibfield  {author} {\bibinfo {author} {\bibfnamefont {R.}\ \bibnamefont
  {Dickman}}, \bibinfo {author} {\bibfnamefont {A.}\ \bibnamefont
  {Vespignani}}, \ and\ \bibinfo {author} {\bibfnamefont {S.}\
  \bibnamefont {Zapperi}},\ }\bibfield  {title} {\bibinfo {title}
  {Self-organized criticality as an absorbing-state phase transition},
  }\href@noop {} {\bibfield  {journal} {\bibinfo  {journal} {Phys. Rev. E}\
  }\textbf {\bibinfo {volume} {57}},\ \bibinfo {pages} {5095} (\bibinfo
  {year} {1998})}\BibitemShut {NoStop}%
\bibitem [{\citenamefont {Vespignani}\ \emph {et~al.}(1998)\citenamefont
  {Vespignani}, \citenamefont {Dickman}, \citenamefont {Mu{\~n}oz},\ and\
  \citenamefont {Zapperi}}]{VDMZ1998}%
  \BibitemOpen
  \bibfield  {author} {\bibinfo {author} {\bibfnamefont {A.}\
  \bibnamefont {Vespignani}}, \bibinfo {author} {\bibfnamefont {R.}\
  \bibnamefont {Dickman}}, \bibinfo {author} {\bibfnamefont {M.~A.}\
  \bibnamefont {Mu{\~n}oz}}, \ and\ \bibinfo {author} {\bibfnamefont {S.}\
  \bibnamefont {Zapperi}},\ }\bibfield  {title} {\bibinfo {title}
  {Driving, Conservation, and Absorbing States in Sandpiles}, }\href@noop {}
  {\bibfield  {journal} {\bibinfo  {journal} {Phys. Rev. Lett.}\ }\textbf
  {\bibinfo {volume} {81}},\ \bibinfo {pages} {5676} (\bibinfo {year}
  {1998})}\BibitemShut {NoStop}%
\bibitem [{\citenamefont {Vespignani}\ \emph {et~al.}(2000)\citenamefont
  {Vespignani}, \citenamefont {Dickman}, \citenamefont {Mu{\~n}oz},\ and\
  \citenamefont {Zapperi}}]{VDMZ2000}%
  \BibitemOpen
  \bibfield  {author} {\bibinfo {author} {\bibfnamefont {A.}\
  \bibnamefont {Vespignani}}, \bibinfo {author} {\bibfnamefont {R.}\
  \bibnamefont {Dickman}}, \bibinfo {author} {\bibfnamefont {M.~A.}\
  \bibnamefont {Mu{\~n}oz}}, \ and\ \bibinfo {author} {\bibfnamefont {S.}\
  \bibnamefont {Zapperi}},\ }\bibfield  {title} {\bibinfo {title}
  {Absorbing-state phase transitions in fixed-energy sandpiles}, }\href@noop
  {} {\bibfield  {journal} {\bibinfo  {journal} {Phys. Rev. E}\ }\textbf
  {\bibinfo {volume} {62}},\ \bibinfo {pages} {4564} (\bibinfo {year}
  {2000})}\BibitemShut {NoStop}%
\bibitem [{\citenamefont {Bak}\ \emph {et~al.}(1987)\citenamefont {Bak},
  \citenamefont {Tang},\ and\ \citenamefont {Wiesenfeld}}]{BTW1987}%
  \BibitemOpen
  \bibfield  {author} {\bibinfo {author} {\bibfnamefont {P.}\ \bibnamefont
  {Bak}}, \bibinfo {author} {\bibfnamefont {C.}\ \bibnamefont {Tang}}, \ and\
  \bibinfo {author} {\bibfnamefont {K.}\ \bibnamefont {Wiesenfeld}},\
  }\bibfield  {title} {\bibinfo {title} {Self-Organized Criticality:
  An Explanation of the $1/f$ Noise}, }\href@noop {} {\bibfield  {journal}
  {\bibinfo  {journal} {Phys. Rev. Lett.}\ }\textbf {\bibinfo {volume} {59}},\
  \bibinfo {pages} {381} (\bibinfo {year} {1987})}\BibitemShut {NoStop}%
\bibitem [{\citenamefont {Dhar}(1990)}]{D1990}%
  \BibitemOpen
  \bibfield  {author} {\bibinfo {author} {\bibfnamefont {D.}\ \bibnamefont
  {Dhar}},\ }\bibfield  {title} {\bibinfo {title} {Self-Organized
  Critical State of Sandpile Automaton Models}, }\href {\doibase
  10.1103/PhysRevLett.64.1613} {\bibfield  {journal} {\bibinfo  {journal}
  {Phys. Rev. Lett.}\ }\textbf {\bibinfo {volume} {64}},\ \bibinfo {pages}
  {1613} (\bibinfo {year} {1990})}\BibitemShut {NoStop}%
\bibitem [{\citenamefont {Manna}(1991)}]{M1991}%
  \BibitemOpen
  \bibfield  {author} {\bibinfo {author} {\bibfnamefont {S.~S.}\ \bibnamefont
  {Manna}},\ }\bibfield  {title} {\bibinfo {title} {Two-state model
  of self-organized criticality}, }\href@noop {} {\bibfield  {journal}
  {\bibinfo  {journal} {J. Phys. A}\ }\textbf {\bibinfo {volume} {24}},\
  \bibinfo {pages} {L363} (\bibinfo {year} {1991})}\BibitemShut
  {NoStop}%
\bibitem [{\citenamefont {Ivashkevich}\ and\ \citenamefont
  {Priezzhev}(1998)}]{IP1998}%
  \BibitemOpen
  \bibfield  {author} {\bibinfo {author} {\bibfnamefont {E.~V.}\ \bibnamefont
  {Ivashkevich}}\ and\ \bibinfo {author} {\bibfnamefont {V.~B.}\ \bibnamefont
  {Priezzhev}},\ }\bibfield  {title} {\bibinfo {title} {Introduction
  to the sandpile model}, }\href {\doibase DOI:
  10.1016/S0378-4371(98)00012-0} {\bibfield  {journal} {\bibinfo  {journal}
  {Physica A}\ }\textbf {\bibinfo {volume} {254}},\ \bibinfo {pages} {97} (\bibinfo {year} {1998})}\BibitemShut {NoStop}%
\bibitem [{\citenamefont {Jensen}(1998)}]{J1998book}%
  \BibitemOpen
  \bibfield  {author} {\bibinfo {author} {\bibfnamefont {H.~J.}\ \bibnamefont
  {Jensen}},\ }\href@noop {} {\emph {\bibinfo {title} {Self-Organized
  Criticality}}}\ (\bibinfo  {publisher} {Cambridge University Press},\ \bibinfo
  {address} {Cambridge, UK},\ \bibinfo {year} {1998})\BibitemShut {NoStop}%
\bibitem [{\citenamefont {Pruessner}(2012)}]{P2012book}%
  \BibitemOpen
  \bibfield  {author} {\bibinfo {author} {\bibfnamefont {G.}\ \bibnamefont
  {Pruessner}},\ }\href@noop {} {\emph {\bibinfo {title} {Self-Organized
  Criticality: Theory, Models and Characterisation}}}\ (\bibinfo  {publisher}
  {Cambridge University Press},\ \bibinfo {address} {Cambridge, UK},\ \bibinfo
  {year} {2012})\BibitemShut {NoStop}%
\bibitem [{\citenamefont {Hinrichsen}(2000)}]{H2000}%
  \BibitemOpen
  \bibfield  {author} {\bibinfo {author} {\bibfnamefont {H.}~\bibnamefont
  {Hinrichsen}},\ }\bibfield  {title} {\bibinfo {title}
  {Non-equilibrium critical phenomena and phase transitions into absorbing
  states}, }\href@noop {} {\bibfield  {journal} {\bibinfo  {journal} {Adv.
  Phys.}\ }\textbf {\bibinfo {volume} {49}},\ \bibinfo {pages} {815} (\bibinfo
  {year} {2000})}\BibitemShut {NoStop}%
\bibitem [{\citenamefont {\'{O}dor}(2004)}]{O2004}%
  \BibitemOpen
  \bibfield  {author} {\bibinfo {author} {\bibfnamefont {G.}\
  \bibnamefont {\'{O}dor}},\ }\bibfield  {title} {\bibinfo {title}
  {Universality classes in nonequilibrium lattice systems}, }\href {\doibase
  10.1103/RevModPhys.76.663} {\bibfield  {journal} {\bibinfo  {journal} {Rev.
  Mod. Phys.}\ }\textbf {\bibinfo {volume} {76}},\ \bibinfo {eid} {663}
  (\bibinfo {year} {2004})}\BibitemShut {NoStop}%
\bibitem [{\citenamefont {Fey}\ \emph {et~al.}(2010{\natexlab{a}})\citenamefont
  {Fey}, \citenamefont {Levine},\ and\ \citenamefont {Wilson}}]{FLW2010}%
  \BibitemOpen
  \bibfield  {author} {\bibinfo {author} {\bibfnamefont {A.}\ \bibnamefont
  {Fey}}, \bibinfo {author} {\bibfnamefont {L.}\ \bibnamefont {Levine}}, \
  and\ \bibinfo {author} {\bibfnamefont {D.~B.}\ \bibnamefont {Wilson}},\
  }\bibfield  {title} {\bibinfo {title} {Driving Sandpiles to
  Criticality and Beyond}, }\href@noop {} {\bibfield  {journal} {\bibinfo
  {journal} {Phys. Rev. Lett.}\ }\textbf {\bibinfo {volume} {104}},\ \bibinfo
  {pages} {145703} (\bibinfo {year} {2010}{\natexlab{a}})}\BibitemShut
  {NoStop}%
\bibitem [{\citenamefont {van Kampen}(1992)}]{vK1992}%
  \BibitemOpen
  \bibfield  {author} {\bibinfo {author} {\bibfnamefont {N.~G.}\ \bibnamefont
  {van Kampen}},\ }\href@noop {} {\emph {\bibinfo {title} {Stochastic Processes
  in Physics and Chemistry}}},\ \bibinfo {edition} {2nd}\ ed.\ (\bibinfo
  {publisher} {North-Holland},\ \bibinfo {address} {Amsterdam},\ \bibinfo
  {year} {1992})\BibitemShut {NoStop}%
\bibitem [{\citenamefont {Harris}(1974)}]{H1974}%
  \BibitemOpen
  \bibfield  {author} {\bibinfo {author} {\bibfnamefont {T.~E.}\ \bibnamefont
  {Harris}},\ }\bibfield  {title} {\bibinfo {title} {Contact
  interactions on a lattice}, }\href@noop {} {\bibfield  {journal} {\bibinfo
  {journal} {Ann. Probab.}\ }\textbf {\bibinfo {volume} {2}},\ \bibinfo {pages}
  {969} (\bibinfo {year} {1974})}\BibitemShut {NoStop}%
\bibitem [{\citenamefont {Dhar}(2006)}]{Dhar2006}%
  \BibitemOpen
  \bibfield  {author} {\bibinfo {author} {\bibfnamefont {D.}\ \bibnamefont
  {Dhar}},\ }\bibfield  {title} {\bibinfo {title} {Theoretical
  studies of self-organized criticality}, }\href@noop {} {\bibfield
  {journal} {\bibinfo  {journal} {Physica A}\ }\textbf {\bibinfo {volume}
  {369}},\ \bibinfo {pages} {29} (\bibinfo {year} {2006})}\BibitemShut
  {NoStop}%
\bibitem [{\citenamefont {Dall'Asta}(2006)}]{D2006}%
  \BibitemOpen
  \bibfield  {author} {\bibinfo {author} {\bibfnamefont {L.}~\bibnamefont
  {Dall'Asta}},\ }\bibfield  {title} {\bibinfo {title} {Exact
  Solution of the One-Dimensional Deterministic Fixed-Energy Sandpile Model},
  }\href@noop {} {\bibfield  {journal} {\bibinfo  {journal} {Phys. Rev. Lett.}\
  }\textbf {\bibinfo {volume} {96}},\ \bibinfo {pages} {058003} (\bibinfo
  {year} {2006})}\BibitemShut {NoStop}%
\bibitem [{\citenamefont {Fey}\ \emph {et~al.}(2010{\natexlab{b}})\citenamefont
  {Fey}, \citenamefont {Levine},\ and\ \citenamefont {Peres}}]{FLP2010}%
  \BibitemOpen
  \bibfield  {author} {\bibinfo {author} {\bibfnamefont {A.}\ \bibnamefont
  {Fey}}, \bibinfo {author} {\bibfnamefont {L.}\ \bibnamefont {Levine}}, \
  and\ \bibinfo {author} {\bibfnamefont {Y.}\ \bibnamefont {Peres}},\
  }\bibfield  {title} {\bibinfo {title} {{Growth rates and explosions
  in sandpiles}}, }\href@noop {} {\bibfield  {journal} {\bibinfo  {journal}
  {J. Stat. Phys.}\ }\textbf {\bibinfo {volume} {138}},\ \bibinfo {pages}
  {143} (\bibinfo {year} {2010}{\natexlab{b}})}\BibitemShut {NoStop}%
\bibitem [{\citenamefont {Park}\ and\ \citenamefont {Park}(2008)}]{PP2008EPJB}%
  \BibitemOpen
  \bibfield  {author} {\bibinfo {author} {\bibfnamefont {S.-C.}\ \bibnamefont
  {Park}}\ and\ \bibinfo {author} {\bibfnamefont {H.}\ \bibnamefont
  {Park}},\ }\bibfield  {title} {\bibinfo {title} {Nonequilibrium
  phase transitions into absorbing states}, }\href@noop {} {\bibfield
  {journal} {\bibinfo  {journal} {Eur. Phys. J. B}\ }\textbf {\bibinfo {volume}
  {64}},\ \bibinfo {pages} {415} (\bibinfo {year} {2008})}\BibitemShut
  {NoStop}%
\bibitem [{\citenamefont {Stauffer}\ and\ \citenamefont
  {Aharony}(1992)}]{Stauffer1992}%
  \BibitemOpen
  \bibfield  {author} {\bibinfo {author} {\bibfnamefont {D.}\
  \bibnamefont {Stauffer}}\ and\ \bibinfo {author} {\bibfnamefont {A.}\
  \bibnamefont {Aharony}},\ }\href@noop {} {\emph {\bibinfo {title}
  {Introduction to Percolation Theory}}},\ \bibinfo {edition} {2nd}\ ed.\
  (\bibinfo  {publisher} {Taylor \& Francis},\ \bibinfo {address} {London},\
  \bibinfo {year} {1992})\BibitemShut {NoStop}%
\bibitem [{\citenamefont {Mollison}(1977)}]{ABS:M1977}%
  \BibitemOpen
  \bibfield  {author} {\bibinfo {author} {\bibfnamefont {D.}~\bibnamefont
  {Mollison}},\ }\bibfield  {title} {\bibinfo {title} {Spatial
  contact models for ecological and epidemic spread}, }\href@noop {}
  {\bibfield  {journal} {\bibinfo  {journal} {J. R. Statist. Soc. B}\
  }\textbf {\bibinfo {volume} {39}},\ \bibinfo {pages} {283} (\bibinfo
  {year} {1977})}\BibitemShut {NoStop}%
\bibitem [{\citenamefont {Grassberger}(1983)}]{ABS:G1983}%
  \BibitemOpen
  \bibfield  {author} {\bibinfo {author} {\bibfnamefont {P.}~\bibnamefont
  {Grassberger}},\ }\bibfield  {title} {\bibinfo {title} {On the
  critical behavior of the general epidemic process and dynamical
  percolation}, }\href@noop {} {\bibfield  {journal} {\bibinfo  {journal}
  {Math. Biosci.}\ }\textbf {\bibinfo {volume} {63}},\ \bibinfo {pages}
  {157} (\bibinfo {year} {1983})}\BibitemShut {NoStop}%
\bibitem [{\citenamefont {Mu\~noz}\ \emph {et~al.}(1999)\citenamefont
  {Mu\~noz}, \citenamefont {Dickman}, \citenamefont {Vespignani},\ and\
  \citenamefont {Zapperi}}]{MDVZ1999}%
  \BibitemOpen
  \bibfield  {author} {\bibinfo {author} {\bibfnamefont {M.~A.}\
  \bibnamefont {Mu\~noz}}, \bibinfo {author} {\bibfnamefont {R.}\
  \bibnamefont {Dickman}}, \bibinfo {author} {\bibfnamefont {A.}\
  \bibnamefont {Vespignani}}, \ and\ \bibinfo {author} {\bibfnamefont
  {S.}\ \bibnamefont {Zapperi}},\ }\bibfield  {title} {\bibinfo
  {title} {Avalanche and spreading exponents in systems with absorbing
  states}, }\href {\doibase 10.1103/PhysRevE.59.6175} {\bibfield  {journal}
  {\bibinfo  {journal} {Phys. Rev. E}\ }\textbf {\bibinfo {volume} {59}},\
  \bibinfo {pages} {6175} (\bibinfo {year} {1999})}\BibitemShut {NoStop}%
\bibitem [{\citenamefont {Grassberger}\ and\ \citenamefont {de~la
  Torre}(1979)}]{GT1979}%
  \BibitemOpen
  \bibfield  {author} {\bibinfo {author} {\bibfnamefont {P.}\ \bibnamefont
  {Grassberger}}\ and\ \bibinfo {author} {\bibfnamefont {A.}~\bibnamefont
  {de~la Torre}},\ }\bibfield  {title} {\bibinfo {title} {Reggeon
  field theory (Schl{\"o}gl's first model) on a lattice: Monte carlo
  calculations of critical behavior}, }\href@noop {} {\bibfield  {journal}
  {\bibinfo  {journal} {Ann. Phys.}\ }\textbf {\bibinfo {volume} {122}},\
  \bibinfo {pages} {373} (\bibinfo {year} {1979})}\BibitemShut {NoStop}%
\bibitem [{\citenamefont {Dhar}(private communication)}]{DharPri}%
  \BibitemOpen
  \bibfield  {author} {\bibinfo {author} {\bibfnamefont {D.}\ \bibnamefont
  {Dhar}} }\href@noop {} {} (\bibinfo {year} {private
  communication})\BibitemShut {NoStop}%
\bibitem [{\citenamefont {Dickman}\ \emph {et~al.}(2001)\citenamefont
  {Dickman}, \citenamefont {Alava}, \citenamefont {A.~Mu\~noz}, \citenamefont
  {Peltola}, \citenamefont {Vespignani},\ and\ \citenamefont
  {Zapperi}}]{DAMPVZ2001}%
  \BibitemOpen
  \bibfield  {author} {\bibinfo {author} {\bibfnamefont {R.}\ \bibnamefont
  {Dickman}}, \bibinfo {author} {\bibfnamefont {M.}\ \bibnamefont {Alava}},
  \bibinfo {author} {\bibfnamefont {M.~A.}\ \bibnamefont {Mu\~noz}},
  \bibinfo {author} {\bibfnamefont {J.}\ \bibnamefont {Peltola}}, \bibinfo
  {author} {\bibfnamefont {A.}\ \bibnamefont {Vespignani}}, \ and\
  \bibinfo {author} {\bibfnamefont {S.}\ \bibnamefont {Zapperi}},\
  }\bibfield  {title} {\bibinfo {title} {Critical behavior of a
  one-dimensional fixed-energy stochastic sandpile}, }\href {\doibase
  10.1103/PhysRevE.64.056104} {\bibfield  {journal} {\bibinfo  {journal} {Phys.
  Rev. E}\ }\textbf {\bibinfo {volume} {64}},\ \bibinfo {pages} {056104}
  (\bibinfo {year} {2001})}\BibitemShut {NoStop}%
\end{thebibliography}%
\end{document}